\UseRawInputEncoding
\documentclass[article,showpacs,groupedaddress,superscriptaddress,nofootinbib,floatfix,preprintnumbers,longbibliography]{revtex4-1}
\usepackage{amssymb,amsmath,graphicx,color}
\usepackage{bm}
\usepackage[tight]{subfigure}
\usepackage[export]{adjustbox}
\usepackage{braket}
\usepackage{amsmath}
\usepackage{physics}
\usepackage{graphicx}
\usepackage{amssymb}
\usepackage[colorlinks=true,citecolor=blue,linkcolor=blue]{hyperref}
\usepackage[table]{xcolor}
\usepackage{cancel}
\usepackage{multirow}

\newcommand{\bbeta}{\mathfrak{b}}
\newcommand{\fstretch}{f_{\textrm{stretch}}}

\newcommand{\sff}{\text{SFF}}
\newcommand{\phia}{\sigma_a}
\usepackage{comment}

\begin{document}

\title{Hydrodynamic Theory of the Connected Spectral Form Factor}

\author{Michael Winer}
\affiliation{Condensed Matter Theory Center and Joint Quantum Institute,
Department of Physics, University of Maryland, College Park, MD 20742, USA}
\author{Brian Swingle}
\affiliation{Brandeis University, Waltham, Massachusetts, USA 02454}

\begin{abstract}
One manifestation of quantum chaos is a random-matrix-like fine-grained energy spectrum. Prior to the inverse level spacing time, random matrix theory predicts a `ramp' of increasing variance in the connected part of the spectral form factor. However, in realistic quantum chaotic systems, the finite time dynamics of the spectral form factor is much richer, with the pure random matrix ramp appearing only at sufficiently late time. In this article, we present a hydrodynamic theory of the connected spectral form factor prior to the inverse level spacing time. We first derive a general formula for the spectral form factor of a system with almost-conserved sectors in terms of return probabilities and spectral form factors within each sector. Next we argue that the theory of fluctuating hydrodynamics can be adapted from the usual Schwinger-Keldysh contour to the periodic time setting needed for the spectral form factor, and we show explicitly that the general formula is recovered in the case of energy diffusion. We also initiate a study of interaction effects in this modified hydrodynamic framework and show how the Thouless time, defined as the time required for the spectral form factor to approach the pure random matrix result, is controlled by the slow hydrodynamics modes. We then extend the formalism to Floquet systems, where a ramp is expected but with a different coefficient, and we derive a crossover formula from the Hamiltonian ramp to the Floquet ramp when the Floquet drive is weak. Taken together, these results establish an effective field theory of chaotic spectral correlations which predicts the random matrix ramp at late time and computes corrections to it at earlier times. 
\end{abstract}
\maketitle
\section{Introduction}

There has been a surge of recent interest~\cite{haake2010quantum,PhysRevLett.52.1,mehta2004random} in the statistics of energy levels of ``chaotic'' quantum systems. Quantum chaos in this loose sense is typically invoked when quantizing a classically chaotic system and in the context of quantum systems that effectively thermalize. It is widely believed that ensembles of such chaotic systems have the same Hamiltonian spectral statistics as ensembles of Gaussian Hermitian random matrices, with examples from nuclear systems~\cite{doi:10.1063/1.1703775,wigner1959group} to condensed matter systems~\cite{bohigas1984chaotic,dubertrand2016spectral,PhysRevLett.121.264101} to holographic theories~\cite{Cotler2017,saad2019semiclassical}. One line of thought even proposes to take random-matrix-like spectral statistics as a \textit{definition} of quantum chaos, which then denotes a super-class of physical systems that includes many-body systems like nuclear matter (which were Wigner's motivation for introducing the random matrix ansatz) and quantized few-body systems (where the connection to classical chaos is clear). From this point of view, classical chaos still has an important relation to quantum chaos, but it is only directly relevant for a subset of all quantum chaotic systems with a recognizable semi-classical limit. 

These ideas also raise a fundamental question: how do we tell which physical systems possess quantum chaos in the spectral sense? Of course, if we have access to the fine-grained energy levels then it is easy to check from the definition, but what if we only have access to more coarse-grained information (as is very often the case in experiment)? For example, thermalization is a widely observed phenomenon that is also closely associated with quantum chaos in the above loose sense, so one can wonder whether chaos in the sense of exhibiting thermalization implies chaos in the sense of random-matrix-like spectral statistics (and vice versa). In this paper, we argue that if an isolated quantum system thermalizes, then it possesses quantum chaos in the spectral sense.

We do this by building an effective description that predicts the apparent universality of random matrix spectral correlations and computes corrections to the pure random matrix answer in the presence of structure in the physical Hamiltonian. The idea of our approach is that corrections to pure random matrix behavior should be controlled by the slow ``hydrodynamic'' modes of the system since these are what distinguish a Hamiltonian with structure from one without. Then, starting from a proper effective theory of these slow modes, we propose a principled modification of that theory which predicts random matrix behavior at late time and can compute corrections to it at earlier times. Although one might have thought that these slow modes only control corrections to pure random matrix behavior, the same effective theory is able to produce the random matrix behavior itself. Thermalization is built into the approach from the start, since we assume that all other modes decay rapidly in time and the effective theory of the slow modes is generic in the sense of effective field theory.

To explain more precisely what our effective theory computes, we first need to specify what we mean by spectral correlations. Suppose $\{ H(J) \}$ is an ensemble of Hamiltonians depending on some random variables (``disorder'') collectively denoted $J$. We say that the ensemble $\{H(J)\}$ is random-matrix-like if the statistical properties of the eigenvalues of $H(J)$ reproduce those of a Gaussian random matrix ensemble with the same symmetry. The spectral statistics of the ensemble can be defined using the density of energy eigenvalues, $\rho(E)$. The simplest object to consider is the ensemble averaged eigenvalue density, $\overline{\rho(E)}$, but as reviewed below, this quantity depends on all the details of the random matrix ensemble and does not provide a universal signature. However, the ensemble averaged pair correlation, $\overline{\rho(E_1) \rho(E_2)}$, is universal across a wide variety of ensembles (even non-Gaussian ones), so it provides a useful diagnostic of randomness in the spectrum.

However, such comparisons between the physical spectra of $\{H(J)\}$ and the predictions of random matrix theory must be applied with care, since a particular chaotic quantum system will typically only have random matrix-like spectral correlations for energy levels that are sufficiently close in energy. This is because physical systems typically have structure in the Hamiltonian, such as constraints arising from spatial locality, which is not present in standard random matrix ensembles and which must be effectively washed out by the dynamics. The effective loss of structure can only happen at long times after the system has come to global equilibrium. For this reason, it is convenient to think about spectral correlations in the time domain by considering the Fourier transform of $\overline{\rho(E_0 + \Delta E/2)\rho(E_0 - \Delta E/2)}$ with respect to the relative energy $\Delta E$. Then longer times correspond to increasingly closely spaced energy levels.

In this paper, we study the Fourier transformed pair correlation, which is known as the spectral form factor (SFF), and present a hydrodynamic theory of the intermediate time spectral properties of generic quantum chaotic systems. By hydrodynamics we mean the recently developed formal effective field theory that governs the slow modes of the system~\cite{Dubovsky_2012,PhysRevD.91.105031,Haehl_2018,crossley2017effective,Jensen_2018}. The relevant modes control the late time behavior of the system as it approaches global equilibrium, and our theory makes a sharp connection between hydrodynamics and the spectral form factor. Moreover, symmetries and hydrodynamics are an inescapable part of the story because time-independent Hamiltonian systems always have at least time translation symmetry and energy conservation, so in a spatially local system there is always at least one slow mode. The hydrodynamic approach we develop predicts that the late time spectral correlations are random-matrix-like and gives quantitative tools to compute corrections to pure random matrix behavior at earlier times. As such, the theory we propose can be viewed as an effective field theory of chaotic spectral correlations.

Before proceeding, we briefly highlight the context of our work. There is a very large literature on quantum chaos extending back many decades. One key paper is \cite{original-thouless} which showed that the variance of the number of single particle energy levels in an energy window was random-matrix-like for energies smaller than the inverse Thouless time defined from particle diffusion. This time originally arose in the context of mesoscopic transport as a measure of the sensitivity of the system to boundary conditions, but it has come to refer to the timescale beyond which quantum dynamics looks random-matrix-like. Other prior investigations of the Thouless time in a many-body setting include \cite{Schiulaz_2019,Gharibyan_2018,Friedman_2019,Altland_2018,Nosaka_2018}, with one result being that the Thouless time can depend on the observable used to define it. There are also a growing number of exact diagonalization studies and analytic results on many-body spectral statistics and spectral form factors including~\cite{PhysRevLett.121.264101,saad2019semiclassical,Garc_a_Garc_a_2017,altland2020late}. \cite{2020Prosen} derives a similar formula to our \ref{eq:GenFormula}. One useful recent review on various aspects of quantum chaos is \cite{D_Alessio_2016}.

To setup a more precise statement of our main results, we first review the basics of random matrix theory (RMT) and the observables of interest. After this short review, we outline our results at the end of the introduction and give a guide to the paper.

\subsection{Random matrix theory and spectral form factor }

A ``single trace'' random matrix ensemble is characterized by two pieces of data. The first datum is the type of matrix (orthogonal, unitary, symplectic) and corresponding Dyson index $\bbeta=1,2,4$. In physical terms, this relates to the number and nature of antiunitary symmetries. The second datum is a potential $V(E)$, where we choose matrix $H$ with probability $dP\propto \prod_{ij}dH_{ij} \exp \left( -\tr V(H)\right)$. These data give a joint probability for the eigenvalues $\{E_i\}$ equal to 
\begin{equation}dP=\frac 1 {\mathcal Z}\prod_{i<j} |E_i-E_j|^{\bbeta} \prod_i e^{-V(E_i)},
\label{eq:RMTpdf}
\end{equation}
where $\mathcal{Z}$ is a normalization. 

This probability distribution can be conveniently interpreted in terms of a ``Coulomb gas'' of eigenvalues as follows. Eq.~\ref{eq:RMTpdf} has the form of a Boltzmann distribution at unit temperature for a gas of 1d particles at positions $E_i$ with logarithmic Coulomb interactions $U_{ij}=\bbeta \log |E_i-E_j|$ subject to an external potential $V$~\cite{mehta2004random}. In this way of thinking, the correlations of the density of particles/eigenvalues,
\begin{equation}
\rho(E)=\sum_{i}\delta(E-E_i),
\end{equation}
form a natural set of observables. The most basic of these observables is the density of states, $\overline{\rho(E)}$, where the overline denotes an average over the random matrix ensemble. For example, in a Gaussian random matrix ensemble in which the potential $V$ is quadratic, this average is well approximated by the famous Wigner semi-circle law. However, the average density of states is non-universal since it depends on the potential $V$. The simplest observable that probes spectral correlations and which does not depend strongly on $V$ is the 2-point function of the density, $\overline{\rho(E_1) \rho(E_2)}$. 

It is common~\cite{heusler2004universal,brezin1997spectral} to package this 2-point function into an object in the time domain called a spectral form factor (SFF), defined here to include a filter function $f$,
\begin{equation}
\sff(T,f)=\overline{|\tr f(H)e^{-iHT}|^2}=\overline{\sum_{i,j}f(E_i)f(E_j)e^{i(E_i-E_j)T}}.
\label{eq:SFFdef}
\end{equation}
Very often we choose $f(H)=\exp(-\beta H)$, which we call the SFF at inverse temperature $\beta$. (In this paper $\bbeta$ is the Dyson index and $\beta$ is inverse temperature. $T$ is time and never temperature.) Another useful choice for $f$ is a Gaussian function zeroed in on a part of the spectrum of interest. The filtered SFF is then the ensemble average of the squared magnitude of the $T$-component of the Fourier transform of $ f(E)\rho(E)$,
\begin{equation}
   \text{SFF}(T,f) = \overline{ \left| \int_{-\infty}^{\infty} dE f(E) \rho(E) e^{-i E T} \right|^2}.
\end{equation}

The SFF of a random matrix breaks into three regimes. First, a slope region, where Eq.~\ref{eq:SFFdef} is dominated by the disconnected part of the 2-point function of $\rho$. Once the system reaches the Thouless time $t_{\text{Th}}$ when all macroscopic degrees of freedom have relaxed, we reach a new stage.
This second state is the ramp, where the disorder-averaged SFF is linear in $T$. In this regime the SFF is given by
\begin{equation}\sff_{\text{ramp}}=\int dE \frac{T}{\pi \bbeta} f^2(E),
\label{eq:RMTAnswer}
\end{equation} 

\begin{figure}
\centering
\includegraphics[scale=0.6]{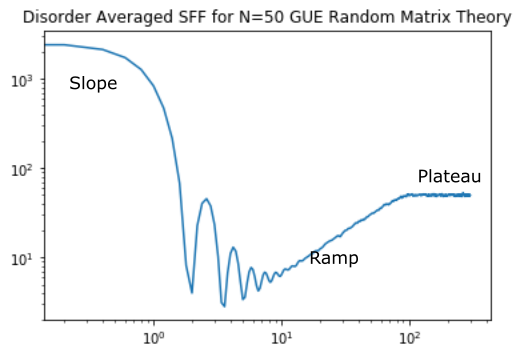}
\caption{The SFF of a simple random matrix system in the Gaussian Unitary Ensemble (GUE) displaying slope, ramp, and plateau behaviors.}
\label{fig:BasicSFF}
\end{figure}

The ramp continues until the Heisenberg time set by the inverse level spacing. At such long times, the off-diagonal terms in equation~\eqref{eq:SFFdef} average to zero and the SFF is a flat plateau. An example log-log plot of a random matrix SFF is shown in Fig.~\ref{fig:BasicSFF}.

\begin{figure}
\centering
\includegraphics[scale=0.6]{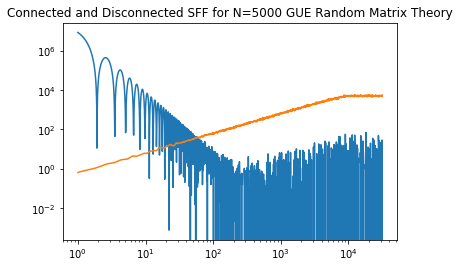}
\caption{The connected (orange) and disconnected (blue) parts of the Spectral Form Factor plotted on a log-log plot. The ramp come and plateau come entirely from the connected bit, and the slope entirely from the disconnected bit. This graph appears to show some small ramp-like behavior for the connected part of the SFF, but that's just because the sample variance (controlled by the connected SFF) is going up so the distribution of values gets wider.}
\label{fig:twoSFF}
\end{figure}

It is further useful to decompose the SFF into connected and disconnected pieces. In terms of the thermodynamic partition function evaluated at imaginary inverse temperature $iT$,
\begin{equation}
    Z(iT,f) = \sum_i f(E_i) e^{-i E_i T},
\end{equation}
the SFF is 
\begin{equation}
    \sff(T,f) = \overline{Z(iT,f) Z^*(T,f)} = \sff_{\text{conn}} + \sff_{\text{disc}},
\end{equation}
where 
\begin{equation}
    \sff_{\text{disc}} = |\overline{Z(iT,f)}|^2
\end{equation}
and
\begin{equation}
    \sff_{\text{conn}} = \sff - \sff_{\text{disc}} = \overline{\left(Z(iT,f)-\overline{Z(iT,f)}\right)\left(Z(iT,f)-\overline{Z(iT,f)}\right)^*}.
\end{equation}
Fig.~\ref{fig:twoSFF} shows the very different behaviors of these two pieces of the SFF. The disconnected part is controlled just by the density of states, so we can more cleanly access the spectral correlations by focusing on the connected part.

\subsection{Overview of results}

Given this background and notation, we can now state our main results. We first fix some terminology used in the paper. The ramp typically refers to the linear in time part of the connected spectral form factor. In a many-body system of $N$ degrees of freedom with no symmetries or slow modes, the ramp is expected to onset after a short relatively short time of order $\log N$.\footnote{This is the time it takes for an exponentially decaying mode of the form $e^{-\lambda t}$ to reach a $1/N$ suppressed amplitude provided the rate $\lambda$ is not $N$-dependent.} The main topic of this article is the modification of the random matrix ramp due to slow modes and non-random matrix features of the system. One could conceivably speak about a `time-dependent ramp coefficient', but we prefer to consider the time period prior to the pure random matrix ramp as distinct regime. In this view, there are four time periods: (1) the very early regime, prior to a time of order $\log N$, when all the details matter, (2) the hydrodynamic regime, when the spectral form factor is determined by the symmetries and slow modes of the system, but is insensitive to other details, (3) the pure random matrix ramp regime, and (4) the plateau regime. 

In this paper we study in detail the connected spectral form factor in the hydrodynamic regime (regime 2) and the pure random matrix ramp regime (regime 3). By contrast, the very early time regime (regime 1) is totally non-universal, and the very late time regime (regime 4) is straightforward to understand microscopically (albeit potentially mysterious from other points of view). We define the Thouless time to be the time it takes for the SFF to come close the pure random matrix ramp, i.e. the crossover time from regime 2 to regime 3. One of our key results is an expression relating the connected SFF to return probabilities (equations \eqref{eq:GenFormula} and \eqref{eq:bigAnswer}), giving precise meaning to the notion that RMT behavior takes over when the system has had time to fully explore Hilbert space~\cite{Schiulaz_2019}.

The remainder of the paper is organized as follows. In Section \ref{sec:nearBlock} we discuss the case of approximate symmetries, corresponding to slowly decaying modes, in general quantum mechanical terms. We show that the connected spectral form factor can be computed in terms of return probabilities for the slow modes. Next, in Section \ref{sec:Hydro} we argue that the theory of fluctuating hydrodynamics, conventionally formulated on the Schwinger-Keldysh contour, can be adapted to the periodic time contour defining the spectral form factor. Focusing on the case of energy diffusion, we show that this ``closed time path'' (CTP) formalism, modified with periodic boundary conditions, recovers the ramp at late time and the return probability formula at intermediate time. Then, in Section \ref{sec:int_effects} we discuss in more detail the effects of interactions and study a partial resummation of perturbative diagrams. Next, in Section \ref{sec:floquet}, we study driven Floquet systems, deriving general formulas for their behavior, rederiving a number of known results, and correctly predicting the crossovers between different regimes of the Floquet drive.

\section{Nearly Block Hamiltonians}
\label{sec:nearBlock}

As discussed above, a particular quantum chaotic system will only approach the random matrix prediction at sufficiently late time. At earlier times, the presence of structure in the Hamiltonian typically implies a significant deviation from the pure random matrix result. In this section, we derive a general formula for the spectral form factor of such systems assuming a certain decomposition into weakly coupled random-matrix-like blocks. This section should be viewed as an introduction to the effects of slow modes on the spectral correlations, one in which the results can be obtained from physically transparent assumptions and elementary manipulations of quantum states. Later in Sec.~\ref{sec:Hydro} we will show how these results are recovered from our hydrodynamic effective theory.

So, suppose the Hamiltonian decomposes into two pieces, $H=H_0+V$, such that $H_0$ breaks into $m$ decoupled blocks and $V$ causes transitions between the blocks. We take the $V$-induced transitions to be slow and the $H_0$ blocks to be random-matrix-like. To compute $Z(iT) = \tr(e^{-i H T})$ (we will add the filter function later), we want to sum over all return amplitudes. Consider a basis for the Hilbert space $|\psi_{(\alpha,i)}\rangle$ labelled by the pair $(\alpha,i)$ where $\alpha$ denotes the block and $i$ indicates a basis vector within a block. Given an initial state $|\psi_{(\alpha,i)}\rangle$, write its time development as
\begin{equation}
    |\psi_{(\alpha,i)}(T) \rangle = \sum_{\beta=1}^{m} \sqrt{p_{\alpha\rightarrow \beta}(T)} |\phi_{\beta,(\alpha,i)}(T) \rangle,
\end{equation}
where $p_{\alpha\rightarrow \beta}(T)$ is the probability to transition to sector $\beta$ after starting in sector $\alpha$ (assumed to be independent of the within-sector label $i$) and $|\phi_{\beta,(\alpha,i)}(T)\rangle$ is the normalized state in sector $\beta$ originating from $|\psi_{(\alpha,i)}\rangle$. The return amplitude is
\begin{equation}
    \langle \psi_{(\alpha,i)}(0) | \psi_{(\alpha,i)}(T)\rangle = \sqrt{p_{\alpha\rightarrow \alpha}(T)} \langle \psi_{(\alpha,i)}(0) | \phi_{\alpha,(\alpha,i)}(T)\rangle.
\end{equation}

The SFF is assembled by summing these amplitudes, taking the squared magnitude, and then averaging over the ensemble. Now, since the dynamics within each sector is random matrix like at the timescales of interest, the diagonal terms should reduce to the within-sector SFF and the off-diagonal terms should be small, 
\begin{equation}
 \sum_{i,j} \overline{\langle \psi_{(\alpha,i)}(0) | \phi_{\alpha,(\alpha,i)}(T)\rangle \langle \psi_{(\beta,j)}(0) | \phi_{\beta,(\beta,j)}(T)\rangle^*} = \delta_{\alpha,\beta} \sff_\alpha(T).
 \label{eq:OffDiagZero}
\end{equation}
Hence, the filtered SFF reduces to
\begin{equation}
    \sff(T,f) = \sum_{\alpha} f(E_\alpha)^2 p_{\alpha\rightarrow \alpha}(t) \sff_\alpha(t).
    \label{eq:GenFormula}
\end{equation}
When the individual blocks are random-matrix-like, then $\sff_\alpha$ is just a linear ramp with a known coefficient, and the evaluation of the $\sff$ reduces to summing over the return probabilities. 

\subsection{Path integral example}

To understand the return probabilities in more detail and to introduce a useful rate-matrix formalism, consider the instructive example of a particle stuck in one of $m$ potential wells, in a kinematic space complicated enough such that the Hamiltonian within each well is well-approximated by a random matrix. The single almost-conserved quantity is an index $a$ ranging from $1$ to $m$.

\begin{figure}
\centering
\includegraphics[scale=0.6]{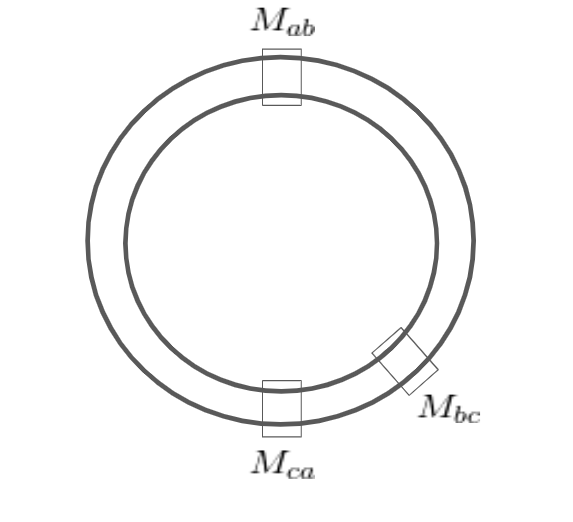}
\caption{The SFF is calculated on a doubled contour for the system. In this configuration, there are three instantons, one taking from well $a$ to well $b$, one shortly after going from $b$ to $c$, then eventually one taking the system from $c$ back to $a$. In between wells the system is well-described by the dynamics within a single sector. }
\label{fig:circleDiagram}
\end{figure}

We can solve this using doubled-system wormhole techniques like those in \cite{saad2019semiclassical}, reviewed in appendix \ref{app:SSS}. 
As a first glimpse of this technology, imagine that the particle dynamics is governed by some classical action such that the trace of the time evolution operator is obtained from a path integral constructed from said action. The SFF is then obtained by doubling this path integral, with one copy for the forward time evolution $e^{-i HT}$ and one copy for the backward time evolution. We will not need the details of this description, just some general properties. In particular, we will not keep track of the detailed dynamics within a well, but we will follow the dynamics of the discrete variable $a$ denoting which well the particle is in. 

Now, the simplest solutions to the equations of motion in a doubled system are ones where $a$ is constant over the entire doubled contour. There are also solutions where $a$ is different on the two contours, but their contributions average to zero because of our assumption that the within-well dynamics is chaotic. Hence, the first rule is that the well index must agree between the two contours. This is analogous to Eq.~\eqref{eq:OffDiagZero}. There are also tunneling events or instantons which take the system from well to well, and we can put all their probabilities into a transition rate matrix $M(E)$. Here $E$ denotes the energy at which the transition is happening. $M(E)$ also has elements on the diagonals to make sure probability is conserved. Note that because these tunneling events happen on a doubled system, the pair of amplitudes, one from each copy of the system, naturally combine to form probabilities. It is these probabilities which are the matrix elements of $M(E)$. An illustration of one configuration which contributes to the path integral is given in figure \ref{fig:circleDiagram}. Note that $M$ is not a Hermitian matrix. It has all negative eigenvalues, except for one zero eigenvalue whose left eigenvector is $(1,1,1...)$ corresponding to conservation of probability). 

To get from the transition matrix to the SFF, the key point is that the same instanton gas that gives us the probability of transfer also shows up in a wormhole-like path integral calculation of the SFF. We start with out in a thermofield double (TFD) for the various approximately disconnected sectors of the Hamiltonian. At each timestep from $t$ to $t+dt$, there is some amplitude (probability from the point of view of a single copy of the system) that the system will go from sector $a$ to sector $b$. This is just $M_{ab}dt$. Multiplying over all timesteps, and requiring that the doubled system start and end in the same sector gives 
\begin{equation}
\textrm{factor from approximate symmetries}=\tr \prod_1^{T/dt} (I+Mdt) =\tr e^{MT}.
\end{equation}
This means that the SFF is given by
\begin{equation}\sff=\int dE \frac{T}{\pi \bbeta} f^2(E) \tr \exp(M(E)T),
\label{eq:bigAnswer}
\end{equation} 
where the $T$ in front still comes from an overall displacement of one side relative to the other. Relative to the pure random matrix result, the coefficient in \eqref{eq:bigAnswer} starts out as $m$ for $m$ wells and goes down to $1$ at long time. It is also worth noting that if there are truly conserved quantities, formula \eqref{eq:bigAnswer} will still give correct results. One interpretation of \eqref{eq:bigAnswer} is as a precise version of the claim that one gets the pure RMT result once enough time has passed for a state to explore all of Hilbert space \cite{Schiulaz_2019}.

\begin{figure}
\includegraphics[scale=0.6]{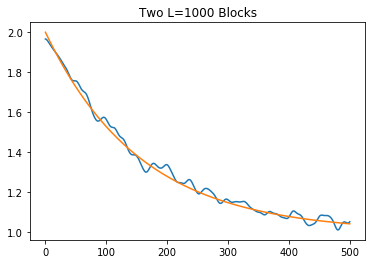}
\includegraphics[scale=0.6]{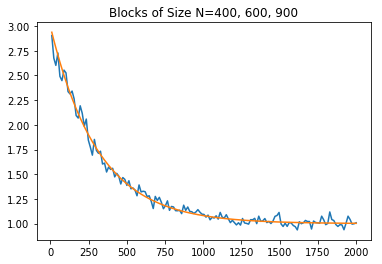}
\includegraphics[scale=0.6]{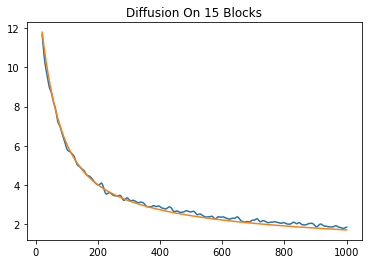}
\caption{Comparison of $\tr e^{MT}$ (orange) vs numerical realization of $\sff/\left(\int dE \frac{T}{\pi \bbeta} f^2(E)\right)$ where $f$ is chosen to be a tightly bunched Gaussian.}
\label{fig:threeGraphs}
\end{figure}

To illustrate the working of formula \eqref{eq:bigAnswer}, suppose we have a random Hermitian matrix of the following form: a $mN\times mN$ complex symmetric matrix, decomposed into $N\times N$ blocks, with elements of variance $J^2$ on the diagonal blocks and variance $k^2J^2$ on the off-diagonal blocks. We can use Fermi's Golden Rule to get transition rates: the squared matrix element is just $k^2J^2$ and the density of states is $\overline{\rho(E)}=\frac{2\sqrt{NJ^2-E^2}}{2\pi J^2}$ (semi-circle law), so the overall rate is $2 k^2\sqrt{NJ^2-E^2}$. Figure \ref{fig:threeGraphs} shows three increasingly complicated scenarios. In the first one, there are two blocks connected with $k=0.04$. In the second, there are three blocks of different sizes. In the third, a chain of blocks where only neighboring blocks are connected. This is analogous to a particle slowly diffusing, where its position is approximately conserved. In each graph, we show the realized ratio of the connected SFF to the predicted single block SFF, and also $\tr e^{M(E)T}$.

\section{Hydrodynamics}
\label{sec:Hydro}

We now turn to the main topic of this paper, the hydrodynamic theory of the connected spectral form factor. Let us quickly recap why hydrodyamics is relevant. As the theory of a system's slow modes, hydrodynamics provides a natural framework in which to evaluate the return probabilities entering the general formula in equations \eqref{eq:GenFormula} and \eqref{eq:bigAnswer}. One might have thought that the late time pure random matrix ramp must still be input by hand, as in the argument in the previous section. However, we will see that the theory developed in this section actually predicts the late time pure random matrix ramp as well. We first motivate the discussion using energy diffusion, then describe our theory in detail in a series of subsections.

\subsection{Energy diffusion and almost conserved quantities}

Energy diffusion is interesting not only as a simple test case, but because it is very generic: any spatially local Hamiltonian system which thermalizes and which does not have additional conserved quantities (the generic case, e.g. due to disorder breaking translation symmetry) is described by this theory at long time/distance. This class of systems fits the previous setting because in the limit of large volume at a fixed time $T$, such a hydrodynamic system has an extensive set of approximate conservation laws. 

As a first step, let us count the number of approximately conserved charges. For linear diffusion, the amplitude of a long-wavelength energy fluctuation with wavevector $k$ decays at rate $D k^2$, where $D$ is the energy diffusion constant. In this case, all modes with wavevector less than $k_T \sim (DT)^{-1/2}$ have not appreciably decayed up to time $T$. In spatial dimension $d$, the number of such modes is 
\begin{equation}
    N_T \sim \sum_k \theta(k_T - |k|) \sim V \int \frac{d^d k}{(2\pi)^d} = \frac{V S_d}{(2\pi)^d}\frac{k_T^d}{d},
\end{equation}
which is extensive in the system size $V$. Hence, since the amplitude of each energy fluctuation with wavevector $|k|<k_T$ is almost conserved, we have an extensive set of almost conserved quantities. The approximately decoupled sectors are then labelled by a choice of amplitude for each mode with $|k|<k_T$. Moreover, within a given sector, all other excitations have decayed, so each sector is plausibly random-matrix-like. Hence, we are in the situation considered in section \ref{sec:nearBlock}.

To go beyond this crude counting of almost conserved modes, we appeal to the formal description of energy diffusion as a problem in fluctuating dissipative hydrodynamics~\cite{Kovtun_2012,Dubovsky_2012,Grozdanov_2015,Endlich_2013}. The particular toolset we use is a modification of the the closed time path (CTP) formalism~\cite{crossley2017effective,glorioso2018lectures}, which itself is a special case of the Schwinger-Keldysh formalism~\cite{Haehl_2017,Kamenev_2009,CHOU19851}. We include a lightning review of this formalism in appendix \ref{app:CTP}, and it is described in great detail in the references. In essence, we couple the conserved energy density and energy current to background fields collectively denoted $A_i$ (with $i$ labelling the forward or backward part of the contour). Suppose the Hamiltonian is modified to $H[A]$ in the presence of background field $A$ (which can depend on time), such that the time evolution is obtained from a time-ordered exponential $U[A]$. Then hydrodynamic correlation functions of the energy density and energy current can be obtained from a generating function of the form $Z[A_1,A_2] = \text{Tr}(U[A_1] \rho U[A_2]^\dagger)$ by differentiating with respect to $A_1$ and $A_2$ and setting $A_1=A_2=0$ at the end of the calculation. 

The CTP formalism is an effective theory of $Z[A_1,A_2]$ in which all fast degrees of freedom have been integrated and only the slow hydrodynamic modes are retained. As reviewed in appendix \ref{app:CTP}, it is particularly natural to formulate this theory using fields that are symmetric and anti-symmetric between the two contours, i.e. $A_r = \frac{A_1+A_2}{2}$ and $A_a = A_1 - A_2$, instead of $A_1$ and $A_2$. These are called $r$-type/classical and $a$-type/quantum variables, respectively. The key idea is that while $Z[A_1,A_2]$ is a non-local object since we integrate over slow modes, we can write it as a path integral over just the slow modes with a local action built from the slow modes.

In more detail, we can introduce slow variables $\sigma_{1,2}$ and write
\begin{equation}
    Z[A_1,A_2] = \int \mathcal{D} \sigma_1 \mathcal{D} \sigma_2 e^{i S_{\text{hydro}}(B_1,B_2)}
\end{equation}
where $B_i = A_i + \partial \sigma_i$ and $S_{\text{hydro}}$ is a local effective action (distinct from the microscopic action defining the system). Within the CTP formalism, the hydro effective action obeys numerous constraints that allow its form to be deduced from effective field theory reasoning.

To ground the discussion with a concrete example, we continue to focus on energy diffusion, but we emphasize that the connection between hydro theories and the SFF is more general. In the CTP framework, the theory of linear diffusion is given by a Lagrangian of the form~\cite{crossley2017effective}
\begin{equation}
L=-\phia\left(\partial_t\epsilon-D\nabla^2\epsilon\right)+i\beta^{-2}\kappa(\nabla \phia)^2, 
\label{eq:CTPLag}
\end{equation}
where $D$ is the diffusion constant, $\kappa$ is the thermal conductivity, and $\nabla^2$ is the spatial Laplacian. One can also define the specific heat $c=\kappa/D$, in terms of which $\epsilon = c \beta^{-1} \partial_t \sigma_r$. One physical interpretation of the fields $\sigma_{a,r}$ is in terms of maps between physical time and an ``internal fluid time''. 

However, even without appealing to any particular interpretation, the basic mechanics of the hydro action are comprehensible. Ignoring for a moment the $\sigma_a$ quadratic term, $\sigma_a$ plays the role of a Lagrange multiplier enforcing the diffusion equation for the energy density $\epsilon$. The effect of the quadratic term in $\sigma_a$ is to introduce stochastic flucuations in the diffusion equation. This can be seen by uncompleting the square by introducing another fluctuating field with a specially chosen quadratic term. Then $\sigma_a$ is again a Lagrange multiplier, but now it enforces a diffusion equation with a stochastic source that generates fluctuations in the amplitudes of the nearly conserved modes. Such fluctuations must be present due to the dissipative nature of the diffusive decay and the fluctuation-dissipation relation.

\subsection{Connecting fluctuating hydrodynamics to spectral statistics}
\label{subsec:hydro_to_sff}

Now, the straightforward view of the role of the CTP formalism in the computation of SFFs is in terms of the return probability in equations \eqref{eq:GenFormula} and \eqref{eq:bigAnswer}. Considering again the example of energy diffusion, we can say that
\begin{equation}
    \tr e^{MT} = \int \mathcal D \epsilon(x,t=0) \Pr(\epsilon(x,t=T)=\epsilon(x,t=0)),
\end{equation}
where $\epsilon(x,t)$ is the energy density at position $x$ and time $t$ and we exclude the spatial zero mode. This can be converted into a path integral over all periodic histories, 
\begin{equation}
    \tr e^{MT} \propto \int \mathcal D \epsilon(x,t)\mathcal D\phia(x,t) e^{iS_{\text{hydro}}[\epsilon,\phia]},
    \label{eq:returnCTP}
\end{equation}
where $\epsilon$ is an $r$-type variable and $\phia$ is the anti-symmetric counterpart of $\epsilon$ in the CTP formalism. In other words, we use the hydro effective action $S_{\text{hydro}}$ to compute the return probabilities. Of course, the zero mode of energy is exactly conserved. In the CTP formalism, it is set by the initial state, but the SFF case, it should be integrated over, weighted by the filter function. The limitation of this point of view is that we seem to be putting in the RMT behavior of individual blocks by hand.

\begin{figure}
    \centering
    \includegraphics[width=.9\textwidth]{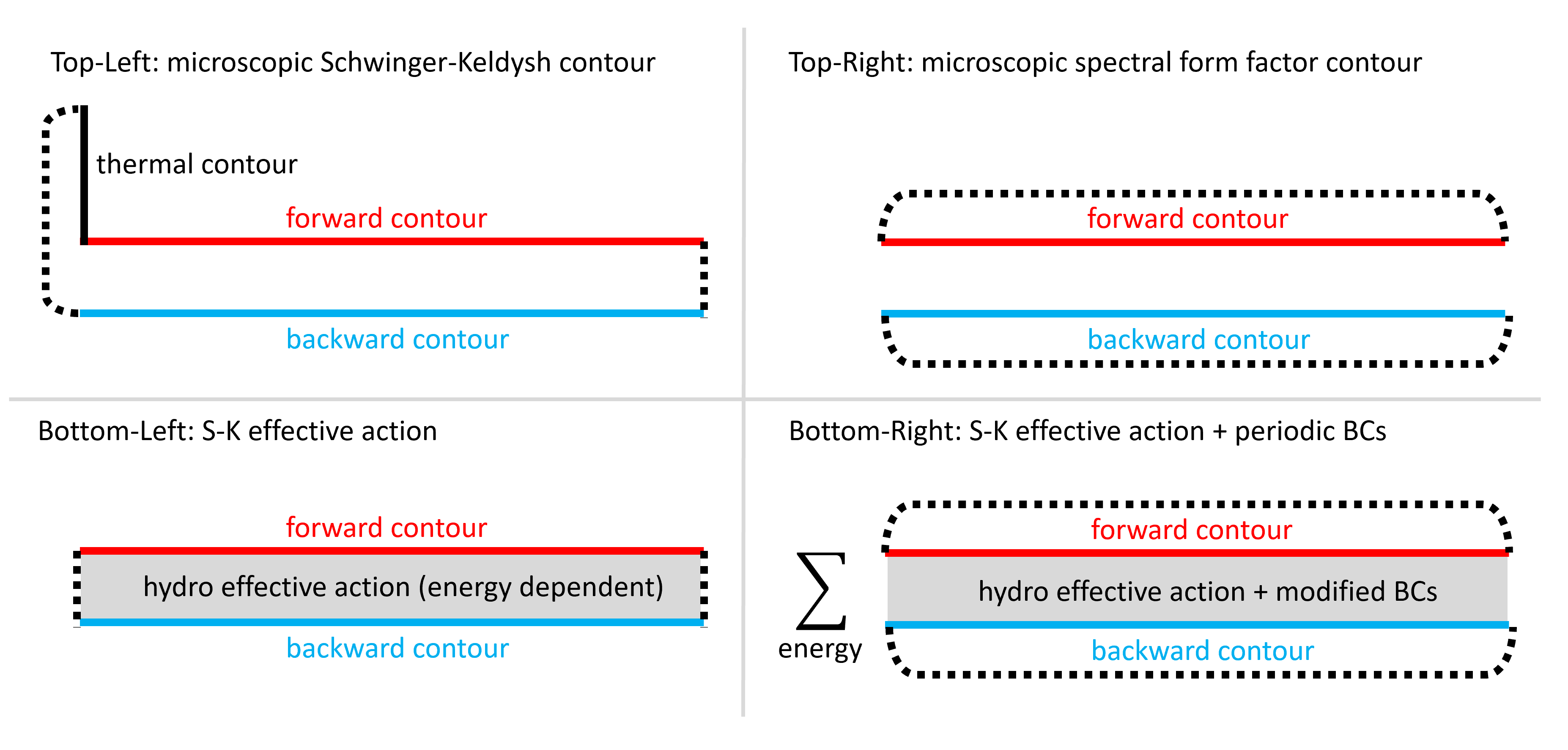}
    \caption{Top-left: The microscopic Schwinger-Keldysh contour which can be used to compute the generating function $Z[A_1,A_2]$ that gives hydrodynamic response functions. Note that the two contours are not explicitly coupled except in the past (from the initial state) and the future (from the trace). Bottom-left: The effective CTP action describing hydrodynamic observables. It can have explicit coupling between the contours, denoted by the grey shading, arising from integrating out fast modes. Top-right: The microscopic SFF contour, again with no explicit coupling between the contours. Bottom-right: Our central hypothesis, that the long-time SFF can be computed using the hydro effective action by modifying the boundary conditions and summing over energies. To rationalize the coupling of contours indicated by the grey shading, there must be an ensemble average which couples the contours.}
    \label{fig:hydro-sff-connection}
\end{figure}

A more general way to look at equation \eqref{eq:returnCTP} is to view it a path integral which will have wormhole-like solutions as in \cite{saad2019semiclassical} (see appendix \ref{app:SSS}). In particular, it is a path integral over two contours going in opposite directions and it focuses on a set of states that (locally) look like an equilibrium thermal state. This connection is summarized in Figure~\ref{fig:hydro-sff-connection}. The key idea is that the contour which defines the generating function of fluctuating hydrodynamics is almost identical to the contour which defines the spectral form factor. They differ only in their boundary conditions in the past and future. In the case of fluctating hydro, there is an initial state and a future trace, and in the case of the SFF, we have periodic boundary conditions and possibly a filter function. Hence, it is natural to suppose that the same hydro effective action can be use to compute both the CTP generating function and the SFF provided we use the appropriate boundary conditions. We call our periodic time modification of the CTP formalism the doubled periodic time (DPT) formalism.

To be completely explicit, here are the assumptions underlying the following analysis of the DPT formalism. Consider a system with `bare' hydrodynamic action $S_{\text{hydro}} = \int d^d x dt L_{\text{hydro}}$ defined on the conventional CTP contour. By bare action we mean that we have integrated out all the fast modes, above some energy scale $\Lambda_{\text{fast}}$, but we have not integrated over any slow modes. Then we assume the following:
\begin{itemize}
    \item First, that the same bare hydro action on the CTP contour can be placed on the SFF contour by simply changing the boundary conditions in time, up to corrections of order $e^{-\Lambda_{\text{fast}}T}$. Physically, the expectation is that the fast modes cannot wrap efficiently around the time circle, and hence the action obtained from integrating them out is not sensitive to the boundary conditions. Note that this statement can only apply to the bare action: once we integrate out modes which can effectively wrap the time circle, then we can get new terms in the action. 
    \item Second, that the bare CTP action with SFF boundary conditions gives the dominant saddle point / phase for the connected SFF for a wide window of time. Specifically, it should be the dominant saddle after times of order $\Lambda_{\text{fast}}^{-1} \log(\text{system size})$ and before the inverse many-body level spacing time. Note that we are relying on the thermodynamic limit to evaluate the SFF by finding the dominant saddle point and computing fluctuations around it.
    \item Third, that there is some averaging over disorder which effectively connects the decoupled SFF contours and rationalizes the interactions between contours in the hydro action. Such averaging is required to make sense of the SFF as a smooth function of time, otherwise one would find an erratic time-dependence. While this disorder average is certainly required, it remains somewhat mysterious from the hydro point of view since the disorder doesn't explicitly appear in the hydro action. Note that the CTP contour already has connectivity between the contours due at least to the future boundary condition, so averaging is not required there if the observables of interest are self-averaging.
\end{itemize}

To the point about boundary conditions, one can object that the boundary conditions shouldn't matter in either case (CTP vs DTP) as far as the fast modes are concerned since they decay rapidly in time. This is almost true, but misses two crucial effects provided by the boundary conditions in the CTP case. The past boundary condition (the initial state) sets the values of conserved quantities. The future boundary condition (the trace) guarantees that, no matter the initial state, the particular fine-grained states that contribute to $Z$ are equal in the far future. This latter condition gives an important fine-grained correlation between the contours. In the DPT theory, we do not have the initial state (although we can include a filter function) or the final trace. This means we need to sum over the values of conserved charges (since these are not fixed and the SFF involves sums over all states) and we need some averaging which produces a similar kind of correlation of fine-grained states on the two contours.

\subsection{Hydrodynamics, Wormholes, and the Thermofield Double}

To begin to flesh out the hydrodynamic theory of the SFF, we first elaborate on the connection to the thermofield double and wormholes. As is pointed out in \cite{saad2019semiclassical} and summarized in appendix \ref{app:SSS}, there are two significant classes of saddle points of a path integral on the SFF contour. One class corresponds to two decoupled saddle points for the two time circles. The other class derives from thermofield double (TFD) solutions, which are correlated between the two contours and which exhibit a free relative time shift and a free total energy. Though the existence of these two classes is general, in the case of holographic systems the TFD solutions also have an interpretation as wormholes. As such, they are literally the ``connected'' part of the SFF. In this context, hydrodynamics appears naturally because it can be viewed as the theory of expanding around a thermofield double solution. Moreover, it is necessary to use hydrodynamics to get a quantitative 1-loop or higher understanding of the size of these contributions. We also caution the reader that for general systems, we must include the fluctuations around the saddle point to get the correct answer (in contrast to large $N$ systems where the saddle point itself is often sufficient).

To calculate the SFF, we need to do a saddle-point expansion for a thermofield double solution on a forward and backwards contour. In this subsection, we discuss the spatial zero modes of the hydrodynamic action. This can be viewed as a theory of zero dimensional systems (such as those with all-to-all interactions) or as the late time limit of a finite-dimensional system in finite volume. The path integral with a quadratic action, in terms of the local relative time shift between the contours $\sigma_a(t)$ and total energy $E_{\text{aux}}$ (the aux notation is chosen to emphasize similarity to \cite{saad2019semiclassical}), is
\begin{equation}
   \sff=\int \frac{\mathcal D E_{\text{aux}}\mathcal D \phia}{2\pi}f^2(E_{\text{aux}})\exp(- i\int dt \sigma_a(t) \partial_t E_{\text{aux}}(t)).
\end{equation} 

The reader unfamiliar with the CTP formalism should think of this as the simplest action that enforces energy conservation, with $\phia(t)$ playing the role of a Lagrange multiplier requiring $\partial_t E_{\text{aux}}=0$. 
Indeed, the integrals over nonzero frequency modes yield delta functions which enforce energy conservation from moment to moment. On the other hand, the integral over the zero frequency modes give exactly the linear ramp, 
\begin{equation}
    \frac{T}{2\pi} \int dE_{\text{aux}}f^2(E_{\text{aux}}).
\end{equation}

It is instructive to compare this answer with the traditional path integral on the CTP contour. In the CTP case, the zero frequency relative time shift is constrained to be zero due to the future boundary condition connecting the contours, but in the DPT case, this relative time shift is naturally unconstrained. Similarly, the total energy integral is weighted by a thermal factor (or the energy distribution of the initial state) in the CTP case, but it is unconstrained (apart from the imposed filter function) in the DPT case.

In the case of a time-reversal invariant Hamiltonian with GOE symmetries, an extra factor of two comes from the possibility of reversing time for one of the contours relative to another, so time $t$ on contour 1 maps to time $-t$ on contour 2. For physical Hamiltonians with GSE symmetries, these cannot be realized without the SFF picking up at least one factor of two in the numerator from degeneracies or blocks, and then we get the GUE answer.

Similar logic can be applied to higher order moments of $Z(iT,f)$ with respect to the disorder average, with the assumption that the relevant saddle points are copies of the dominant DPT saddle point. There are different cases depending on the symmetry. When the symmetry matches the GUE ensemble, then $Z$ is a complex number and the moments of interest are $Z^k Z^{*k}$. There are $k$ forward contours and $k$ backwards contours. Thus there are $k!$ ways to connect the forward and backwards contours into pairs. Once this is done, each one has a free $E_{\text{aux}}$ and a free relative time shift. Thus the $2k$-th moment of $Z$ (assuming there are no additional symmetries) is
\begin{equation}
   \overline{Z^k(T,f)Z^{*k}(T,f)}=k!\left(\frac{T}{2 \pi}\int dE f^2(E)\right)^k.
\end{equation}

In another case, there is an operator $O$ which anticommutes with $H$, and the spectrum has $E\leftrightarrow -E$ symmetry. Provided $f$ is even, $Z$ is always real, and there is no difference between forwards and backwards contours of the SFF. So there are $(2k)!!$ pairings and the answer is
\begin{equation}
    \overline{Z^k(T,f)Z^{*k}(T,f)}=(2k)!!\left(\frac{T}{\bbeta \pi}\int dE f^2(E)\right)^k.
\end{equation}
These are exactly the moments one would get for complex or real Gaussian variables respectively, which is also what one would get from RMT. More surprisingly, we got this without specifying the type of disorder, which indicates that hydrodynamics knows about universal features of disordered systems provided the disorder is not so strong that it changes the structure of the hydro theory. 

\subsection{SFF of Fluctuating Diffusive Hydrodynamics}

Having seen that the hydro theory does in fact predict a linear ramp at late time, we will now evaluate the quantity $\exp(MT)$ for the linear theory of energy diffusion. As we saw above, the ramp comes from the spatial zero modes, and the sum over return probabilities comes from the other spatial modes. Most of the calculation about to be shown is generic for any diffusing substance, but for concreteness we continue to use the language of energy diffusion. In the CTP framework, the theory of linear diffusion is given by a Lagrangian of the form~\cite{crossley2017effective}
\begin{equation}
L=-\phia\left(\partial_t\epsilon-D\nabla^2\epsilon\right)+i\beta^{-2}\kappa(\nabla \phia)^2, 
\label{eq:CTPLag}
\end{equation}
where $D$ is the diffusion constant, $\kappa$ is the thermal conductivity, $\nabla^2$ is the Laplacian. One can also define the specific heat $c=\kappa/D$, in terms of which $\epsilon = c \beta^{-1} \partial_t \sigma_r$. Note that since these physical properties typically vary with temperature/energy density, they should be regarded as functions of the zero mode $E_{\text{aux}}$ and we must integrate the final result over energy. For the analysis in this subsection, we consider the total energy of the system to be fixed and known.

Now, because the action is quadratic, the path integral breaks up into a product over different spatial wavevectors. Hence, the sum over return probabilities is
\begin{equation}
\tr e^{M T}=\prod_k \int d\epsilon_{k,\textrm{init}} p(\epsilon_{k,\textrm{final}}=\epsilon_{k,\textrm{init}})
\label{eq:return}
\end{equation}
Looking at a particular wavevector $k$, let the amplitude at time $t=0$ be $\epsilon_k$. At time $t=T$, the amplitude is given by some probability distribution with mean $e^{- \gamma_k T} \epsilon_k$, where $\gamma_k$ is the decay rate, and variance $\sigma^2(T)$. For the linear theory above, this distribution is a Gaussian,
\begin{equation}
    p(\epsilon_{k,\textrm{final}},T) = \frac{\exp\left( - \frac{(\epsilon_{k,\textrm{final}} - e^{-\gamma_k T} \epsilon_k)^2 }{2 \sigma^2(T)}\right)}{\sqrt{2\pi \sigma^2(T)}},
\end{equation}
although the precise shape turns out not to matter. The return probability integrated over the initial condition is
\begin{equation}
    \int d\epsilon_k p(\epsilon_{k,\textrm{final}}=\epsilon_k,T) = \frac{1}{1-e^{-\gamma_k T}},
\end{equation}
independent of the variance $\sigma^2(T)$.

For the DPT theory above with periodic boundary conditions, $\gamma_k = D k^2$, and the allowed values of $k$ are $k \in (2\pi/L)\mathbb{Z}^d$ where $L$ is the linear size, so that $V=L^d$. This implies that
\begin{equation}
    \tr e^{M T} = \prod_{ k \in (2\pi/L)\mathbb{Z}^d} \frac{1}{1 - e^{- Dk^2 T}}.
\end{equation}
For more general shapes, the decay rates are given by the  eigenvalues $\lambda$ of the Laplacian $\nabla^2$ (which are non-positive). Hence, the general formula is 
\begin{equation}
\tr e^{M T}=  \prod_{\lambda \in \text{spec}(\nabla^2)} \frac{1}{1-e^{D\lambda T}}
\label{eq:diffShape}
\end{equation}

Considering times that are short enough that many modes have not decayed, so that we may ignore the discreteness of the spectrum of $\nabla^2$, the result for a box of volume $V$ is 
\begin{equation}
\log \tr e^{M T} =\frac{V}{(2\pi)^d}\int d^d k \sum_{j=1}^\infty \frac {\exp(-jDk^2T)}j=
\frac{V}{(2\pi)^d}\sum_j \frac 1j \left(\frac \pi {jDT}\right)^{d/2}=V\left(\frac{1}{4\pi DT}\right)^{d/2} \zeta(1+d/2).
\label{eq:diffusionSFF}
\end{equation}
When specialized to one dimension, this agrees exactly with the result in \cite{Friedman_2019} obtained for a particular Floquet model where the diffusing substance was a conserved $U(1)$ charge. At longer times, we see that the slowest modes control the approach to the linear pure random matrix ramp. In particular, when $T$ is large compared to the Thouless time $t_{\text{Th}} \sim V^{2/d}/D$, the trace is exponentially close to unity, $\tr e^{M T} \sim 1 + \mathcal{O}(e^{-T/t_{\text{Th}}})$.

\subsection{Direct evaluation of the SFF path integral}
\label{subsec:direct}

Equation \eqref{eq:diffShape} can also be derived by directly computing the path integral taking into account the periodic boundary conditions in time. Note that the zero mode, $k=0$, requires special attention. It corresponds to the exactly conserved quantity, and the divergence in $(1-e^{-\gamma T})^{-1}$ when $\gamma=0$ should be replaced with a sum over the allowed values of the conserved charge, as follows from the trace formula. This all follows directly from the path integral, as we now show.

The SFF integral is
\begin{equation}
\begin{split}
\sff=\int \mathcal D \epsilon\mathcal D \sigma_a \exp(i S_{\text{hydro}}),\\
S_{\text{hydro}}=\int dVdt\bigg(-\sigma_a(\partial_t-D\Delta)\epsilon+iD\kappa\sigma_a \Delta\sigma_a\bigg).
\end{split}
\label{eq:hydroSFFIntegral}
\end{equation}
If we break the time circle into $ T/\Delta t$ segments, then the measure is 
\begin{equation}
  \mathcal{D} \epsilon \mathcal D \sigma_a = \prod_{x} \prod_{\ell=0}^{T/\Delta t-1} \frac{d \epsilon(x, t = \ell \Delta t) d\sigma_a(x,t = \ell \Delta t)}{2\pi}.
\end{equation}
The $2\pi$s are to enforce proper normalization of delta functions imposing the hydro equations. 

$S_{\text{hydro}}$ is a translation-invariant Gaussian function, so we can break path integral \eqref{eq:hydroSFFIntegral} into a product over spatial modes $k$ with $D\Delta=\lambda_k$, and then over temporal frequencies. These dimensionless frequencies, the eigenvalues of the $ dt \partial_t$ matrix, are the $T/\Delta t$ complex numbers $i\omega$ obeying $(i\omega+1)^{T/\Delta t}=1$. Going to the basis of these modes, we have
\begin{equation}
\begin{split}
\sff=\prod_k \sff_k , \\
\sff_k=\prod_\omega \frac{1}{i\omega-\lambda_k \Delta t}, 
\end{split}
\label{eq:hydroProducts}
\end{equation}
Note that $\sff_k$ is a product over roots of unity of the form $\prod \left[ i\omega +1 - (1+ \lambda_k \Delta t)\right]$. For odd $T/\Delta t$, the result is
\begin{equation}
\sff_k=\left(1-(1+\lambda_k \Delta t)^{T/\Delta t}\right)^{-1} =_{\Delta t \rightarrow 0} \left(1-e^{\lambda_k T}\right)^{-1}.
\end{equation}
After multiplying together the contribution from different momentum modes, we precisely recover the return probability formula. 

The one mode which can't be evaluated this way is the hydrodynamic zero mode corresponding to the total energy, since it has a vanishing action. Instead, one is forced to do the full integral $\int \frac{dE d\sigma_a}{2\pi}$ with $\int d\sigma_a = T$. For systems with time-reversal symmetry there is an additional solution where contour 2 is reversed, so the $2\pi$ in the denominator becomes $\pi$. Thus, the quadratic hydro theory with periodic temporal boundary conditions correctly recovers \eqref{eq:bigAnswer} and does not require the random matrix ramp to be put in by hand.

\subsection{Subdiffusive Hydrodynamics}

As an aside, for some systems, such as fracton systems with multipole conservation~\cite{Gromov_2020,Nandkishore_2019} or systems near a localization transition, one can get subdiffusive dynamics of a conserved density. This can be taken into account by replacing $\nabla^2$ with $\nabla^{2n}$. In this case, the analogue of equation \eqref{eq:diffusionSFF} is
\begin{equation}
\begin{split}
\log \tr e^{M T} =\frac{V}{(2\pi)^d}\int d^d k \sum_{j=1}^\infty \frac {\exp(-jD_nk^{2n}T)}j\\
= \frac{V}{(2\pi)^d}\frac{S_d}{2n}\Gamma\left(\frac{d}{2n}\right) \sum_j \frac 1j \left(\frac 1 {jD_nT}\right)^{d/2n}\\
= V{(2\pi)^d}\frac{S_d}{2n}\Gamma\left(\frac{d}{2n}\right)\left(\frac{1}{D_nT}\right)^{d/2n} \zeta(1+d/2n),
\label{eq:subdiffusionSFF}
\end{split}
\end{equation}
which agrees with the result obtained in \cite{moudgalya2020spectral} for such a system.

\section{Interaction Effects}
\label{sec:int_effects}

In this section, we investigate interaction effects. We make some general comments and exhibit one class of diagrams which can be summed to give a qualitative modification of the diffusive scaling inherent in the quadratic theory of diffusion. We restrict ourselves to a discussion based on perturbation theory. Tp set up a perturbation theory, we must first obtain the quadratic Green's functions. For the DPT formalism, these are obtained from the conventional CTP Green's functions by summing over images. Then one can consider a variety of loop diagrams using vertices from the hydro action and the DPT-modified quadratic Green's functions.

\begin{figure}
    \centering
    \includegraphics[scale=0.5]{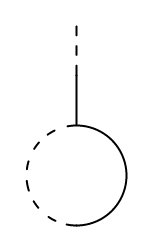}
    \caption{An $arr$-type vertex with an $a$-type variable (dashed line propagator) and an $r$-type variable (solid line propagator) contracted together. In the traditional CTP formulation this would be impossible, but with periodic time there is a contribution from one or more wrappings around the time circle. At long times, contributions from non-trivial wrappings are suppressed by factors of $e^{-T/t_{\text{Th}}}$.}
    \label{fig:tadpole}
\end{figure}

Among the effects present in these diagrams, we expect all the usual effects present in the conventional CTP formalism. These include modifications of various transport coefficients and modifications of the complex structure of various Green's functions arising from loops of the slow modes~\cite{Chen_Lin_2019}. However, in the DPT case, there are qualitatively new effects arising from time periodicity, including classes of diagrams such as the tadpole in Figure~\ref{fig:tadpole} which would identically vanish in the CTP formalism. This section is devoted to a first study of such DPT-specific interaction effects in the context of diffusion.

\subsection{Deformations of the Hamiltonian}

To give a simple warm up, we first discuss how deformations of the Hamiltonian manifest in the DPT formalism. Consider a Hamiltonian $H=H_0 + g  \delta H$. The derivative of $Z(iT) Z(-iT)$ with respect to $g$ is
\begin{equation}
    \delta |Z|^2 = \tr \left( i\delta H e^{iHT} \right)\tr e^{-iHT}-\tr e^{iHT}\tr \left( i\delta He^{-iHT}\right).
\end{equation}
Viewing the two contours as two copies of the system, this expression can be thought of as inserting $i \delta H \otimes I - I \otimes i \delta H$ into the DPT path integral. Because it is anti-symmetric between the two contours, it corresponds to an $a$-type variable in DPT formalism, so the expression for $\delta |Z|^2$ is given by the expectation value of an $a$ variable. In the standard CTP case, such an expectation value would be exactly zero. But in the DPT case, one can get a non-zero result. This had to be so, given our result above, since such a perturbation can certainly change the value of the diffusion constant, and thus the overall answer.

To show how this comes about in the formalism, consider an $arr$-type interaction. This vertex allows for diagrams such as in Figure \ref{fig:tadpole}, which give a nonzero imaginary expectation value to $a$-type variables due to propagators that wrap around the $T$ circle. When $T$ is less than the Thouless time, such wrapping effects are not suppressed and the DPT formalism predicts that the SFF is sensitive to the deformation. However, at times long compared to the Thouless time, the effects of the periodic identification are exponentially small and the formalism predicts that $a$-type variables should have approximately zero average. This observation gives additional insight into how the hydrodynamic DPT formalism encodes the universality of the pure random matrix ramp at late time.

\subsection{Comments on renormalization group approaches}

We next discuss how the conventional scaling analysis of the CTP formalism is modified by time periodicity. To begin, let us recall a simple version of a renormalization group transformation on the hydro Lagrangian for diffusion. Recall that the quadratic Lagrangian is
\begin{equation}
    L_0=-\phia\left(\partial_t\epsilon-D\nabla^2\epsilon\right)+i\beta^{-2}\kappa(\nabla \phia)^2.
\end{equation}
We will focus on the leading perturbations as encoded in 
\begin{equation}
 \Delta L =\frac{\lambda}{2} \nabla^2 \phia\epsilon^2+\frac{\lambda'}{3}\nabla^2 \phia\epsilon^3+ic\beta^{-2} (\nabla \phia)^2(\tilde \lambda \epsilon+\tilde\lambda' \epsilon^2).
\end{equation}

We want a scaling transformation which leaves this quadratic action invariant. From the $\sigma_a \nabla^2 \epsilon$ and $(\nabla \sigma_a)^2$ terms, we see that $\sigma_a$ and $\epsilon$ should be taken to have the same scaling dimension. Similarly, comparing the $\sigma_a \partial_t \epsilon$ and $\sigma_a \nabla^2 \epsilon$ terms, we see that time and space should scale with relative power of two. Under a rescaling
\begin{equation}
    x \rightarrow \lambda x ,
\end{equation}
\begin{equation}
    t \rightarrow \lambda^2 t,
\end{equation}
\begin{equation}
    (\sigma_a, \epsilon) \rightarrow \lambda^{-\Delta_0} (\sigma_a, \epsilon), 
\end{equation}
the quadratic action goes into
\begin{equation}
    I = \int  d^d x dt L_0 \rightarrow \int d^d x dt \lambda^{d+2} \lambda^{-2 - 2 \Delta_0} L_0.
\end{equation}
Hence, the quadratic action is invariant if $\Delta_0 = d/2$.

With this scaling, the dimensions of the non-quadratic operators in $\Delta L$ are
\begin{equation}
    2 + 3 \Delta_0, 2 + 4 \Delta_0.
\end{equation}
All these operators are irrelevant since
\begin{equation}
    2 + 4 \Delta_0 > 2 + 3 \Delta_0 > d+2.
\end{equation}
Of course, these irrelevant operators can still have important effects, but the theory is weakly coupled at low energies in this RG sense.

Now, how is this picture modified by time periodicity? When formulating an RG of the DPT path integral, we expect some general features:
\begin{itemize}
    \item The system size $L$ and time $T$ will flow under the RG. When using diffusive scaling to define the flow, the combination $L^2/T$ will be preserved under the RG.
    \item The bare action is defined with a cutoff in momentum of $\Lambda_k \sim 1/\ell_0$ and a cutoff in frequency of $\Lambda_\omega \sim 1/\tau_0$. Here $\ell_0$ and $\tau_0$ are some sort of mean free path and scattering time, respectively.
    \item The bare action with this cutoff should be identical to the CTP bare action up to corrections of the form $e^{-T/\tau_0}$, since we've only integrated out modes with frequencies higher than $\Lambda_\omega$.
    \item Upon integrating out additional modes at frequency $\omega$, we expect any new terms in the action to respect CTP rules up to $e^{-T \omega}$ corrections since these modes cannot effectively wrap the time circle. This breaks down if we reach a point where $\omega T \sim 1$.
\end{itemize}
These expectations are in essence an elaboration of the arguments made in Section \ref{subsec:hydro_to_sff} for use of the CTP bulk Lagrangian on the SFF contour.

\subsection{A First Feynman Diagram}
\label{subsec:figureEight}

We now compute some novel effects of interactions that arise due to time periodicity in an interacting version of~\eqref{eq:CTPLag}. We focus on simple 1-loop effects in this subsection. Following the conventions in \cite{Chen_Lin_2019} and \cite{crossley2017effective}, we consider the Lagrangian
\begin{equation}
L = i\beta^{-2}\kappa(\nabla \phia)^2-\phia\left(\partial_t\epsilon-D\nabla^2\epsilon\right)
+\frac{\lambda}{2}\nabla^2 \phia\epsilon^2+\frac{\lambda'}{3}\nabla^2 \phia\epsilon^3+ic\beta^{-2} (\nabla \phia)^2(\tilde \lambda \epsilon+\tilde\lambda' \epsilon^2).
\end{equation}
These interactions arise from a variety of sources, including the fact that the parameters of the effective theory depend on the background density. Hence, local fluctuations in the density will induce variations in the parameters of the Lagrangian. Such effects can be captured by including non-Gaussian terms in the effective action.

\begin{figure}
\centering
\includegraphics[scale=0.5]{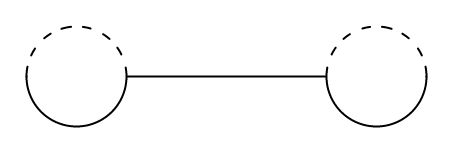}
\caption{Two $\lambda$ vertices. The propagator along the dumbbell is infinite, and the diagram diverges unless $\lambda=0$ and thus we are at a local extremum of diffusivity.}
\label{fig:dumbbell}
\end{figure}

Our interest is in the new classes of diagrams allowed by periodic time. One interesting diagram is the dumbbell in Figure \ref{fig:dumbbell}. This diagram diverges unless $\lambda=0$. This condition is equivalent to requiring we expand around an energy density which is an extremum of diffusivity. This makes perfect sense in light of the SFF formula in equation \eqref{eq:diffusionSFF}, which suggests that the dominant contribution should come from the minimum of diffusivity. If instead we add a filter function which localizes the total energy integral around some $\bar E$, then this effectively adds a mass to the zero mode and removes the divergence.

\begin{figure}
\centering
\includegraphics[scale=0.5]{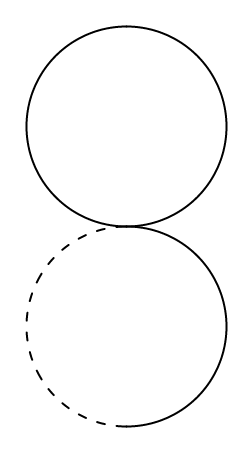}
\caption{A diagram contributing to the zero-point energy. In the case of normal hydrodynamics, this contribution would vanish since the $ra$ propagator would vanish at zero time difference.}
\label{fig:figureEight}
\end{figure}

Suppose we consider an extremum of the transport coefficients (e.g. the case of charge diffusion at half filling). Then the dumbbell diagram is set to zero, and the leading order corrections to the action simplify to
\begin{equation}
    \Delta \mathcal{L} =  \frac{\lambda'}{3} \nabla^2 \sigma_a \epsilon^3 + i c\beta^2\tilde{\lambda}'  (\nabla \sigma_a)^2 \epsilon^2.
\end{equation}
We have the propagators
\begin{equation}
    G_{ar}(k,t) = i \frac{e^{Dk^2(t-T)}}{1-e^{-Dk^2 T}},
\end{equation}
\begin{equation}
    G_{ra}(k,t) = i \frac{e^{-Dk^2t}}{1-e^{-Dk^2 T}},
\end{equation}
and
\begin{equation}
    G_{rr}(k,t) = \frac{\beta^{-2}\kappa}{D} \left( \frac{e^{-Dk^2 t}}{1-e^{-Dk^2 T}} + \frac{e^{-Dk^2 (T-t)}}{1-e^{-Dk^2 T}} \right).
\end{equation}

The leading perturbative correction to the DPT path integral requires that we evaluate the expectation value of $\Delta S = \int \Delta \mathcal{L}$ with respect to the quadratic theory, with path integral $Z_{\text{DPT},0}$. This corresponds to the diagram in Figure \ref{fig:figureEight}. The DPT path integral is then approximately 
\begin{equation}
    Z_{\text{DPT}} \approx Z_{\text{DPT},0} e^{i \langle \Delta S \rangle_0}.
\end{equation}
The expectation of $\Delta I$ contains two terms, one proportional to $\lambda'$ and one proportional to $\tilde \lambda'$. The leading $\tilde \lambda'$ vanishes by symmetry ($\nabla G_{ra}(r=0,0) = 0$). The $\lambda'$ term gives
\begin{equation}
    \langle i \Delta S \rangle_0 = i \lambda' V T [\nabla^2 G_{ra}](0,0) G_{rr}(0,0).
\end{equation}
With the convention that $\phia$ appears slightly ahead of $\rho$ in the action, the two Green functions are
\begin{equation}
    [\nabla^2 G_{ra}](0,0) = \frac{1}{V} \sum_k (-k^2)i \frac{e^{-Dk^2 T}}{1-e^{-Dk^2 T}}
\end{equation}
and
\begin{equation}
    G_{rr}(0,0) = \frac{1}{V} \sum_k \frac{\beta^{-2} \kappa}{D} \frac{1+ e^{-Dk^2 T}}{1-e^{-Dk^2T}}.
\end{equation}
We exclude $k=0$ from these sums which corresponds to fixing the total charge, e.g. using a filter function. In the continuous wavevector regime, the contribution is
\begin{equation}
     \langle i \Delta S \rangle_0 = i \lambda' V T \left( - i \frac{1}{(2\pi)^d} \frac{1}{2DT} \left( \frac{\pi}{DT}\right)^{d/2} \zeta(1+d/2) \right) \left( \frac{\Omega_d \Lambda_k^d}{d (2\pi)^d} + 2 \frac{1}{(2\pi)^d} \left( \frac{\pi}{DT}\right)^{d/2} \zeta(d/2) \right),
\end{equation}
where $\Lambda_k$ is a momentum cutoff. 

To interpret this result, note that the quadratic term (still in the continuous wavevector regime) can be written solely as a function of the scaling variable $u=DT/L^2$ where $V=L^d$, 
\begin{equation}
    Z_{\text{DPT},0} = \exp\left(\frac{c_1}{u^{d/2}}\right).
\end{equation}
However, the correction term violates this scaling collapse since
\begin{equation}
    e^{i \langle \Delta S \rangle_0} = \exp \left( \frac{c_2}{u^{d/2}} + \frac{c_3}{V u^d} \right),
\end{equation}
but at fixed $u$ in the large $V$ limit, the second term goes to zero. To get stronger effects, one has to consider the resummations discussed in the context of energy diffusion.
\subsection{Interaction Effects and Resummation}
Still considering Lagrangian
\begin{equation}
L = i\beta^{-2}\kappa(\nabla \phia)^2-\phia\left(\partial_t\epsilon-D\nabla^2\epsilon\right)
+\frac{\lambda}{2}\nabla^2 \phia\epsilon^2+\frac{\lambda'}{3}\nabla^2 \phia\epsilon^3+ic\beta^{-2} (\nabla \phia)^2(\tilde \lambda \epsilon+\tilde\lambda' \epsilon^2),
\end{equation}
we now perform a more sophisticated analysis than in subsection \ref{subsec:figureEight}. The behavior of the $rr$ propagator should be modified by the existance of the $rr$ self-energy

\begin{figure}
\centering
\includegraphics[scale=0.5]{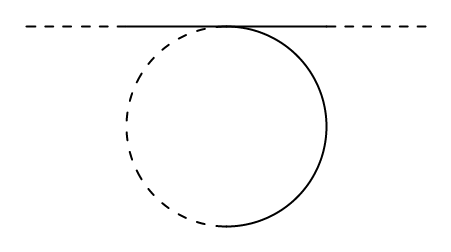}
\caption{A diagram contributing to the $rr$ self energy. In the case of normal hydrodynamics, there would be no $rr$ self energy.}
\label{fig:selfEn}
\end{figure}

This self energy comes from diagrams like the one in figure \ref{fig:selfEn}.
This diagram would normally vanish in the CTP setup, but here it contributes a non-vanishing $rr$ self energy. We treat this term self-consistently by adding an undetermined self energy to the action and fixing it self-consistently. Since $\phia$ self-interactions still need to have a derivative in front of them by CTP rules, a constant $rr$ self energy $\Sigma$ is indeed the most IR-important term we can add. Then the propagators are
\begin{equation}
\begin{split}
G_{rr}(\omega,k)=\frac{\beta^{-2}\kappa k^2}{\Sigma D\kappa k^2+(D^2k^4+\omega^2)}\\
G_{ra}(\omega,k)=\frac{i\omega+Dk^2}{\Sigma D\kappa k^2+(D^2k^4+\omega^2)}\\
G_{aa}(\omega,k)=\frac{\Sigma}{\Sigma D\kappa k^2+(D^2k^4+\omega^2)}.
\end{split}
\end{equation}  
To solve for the self energy, we will need to sum $G_{ra}$ over all frequencies $\omega=2\pi n/T$. The sum is 
\begin{equation}
\sum_{\omega=2\pi n/T} G_{ra}(\omega,k)=\frac{Dk^2}{\sqrt{D^2k^4+\Sigma D\kappa k^2}}\frac{\exp(-\sqrt{D^2k^4+\Sigma D\kappa k^2}T)}{1-\exp(-\sqrt{D^2k^4+\Sigma D\kappa k^2}T)},
\end{equation}
which gives an expression for $\Sigma$:
\begin{equation}
\Sigma=\lambda' \sum_n\int \frac{d^d k}{(2\pi)^d} k^2 \frac{Dk^2}{\sqrt{D^2k^4+\Sigma D\kappa k^2}}\exp(-n\sqrt{D^2k^4+\Sigma D\kappa k^2}T)
\end{equation}
There isn't an IR divergence on the right hand side, so to leading order in large $T$, the $\Sigma$ dependence on the right hand side can be dropped. We are left with
\begin{equation}
\Sigma=\lambda' \sum_n\int \frac{d^d k}{(2\pi)^d} k^2 \exp(-nDk^2T)=\lambda'\frac{d}{2DT} \sqrt{\frac{1}{4\pi DT}}^d\zeta(1+d/2),
\end{equation}
where for the last equality we work in the time regime where the wavevector may be treated as continuous. This result also has an intuitive interpretation. At a minimum of diffusivity, the result \eqref{eq:diffusionSFF} gets a quadratic dependence on $\epsilon$ which is exactly the self-energy.

When in this self-energy term important? Ignoring factors of $\kappa$ and $D$ (which are effectively unit conversions), the two terms in the square root are comparable when $k \sim k_\Sigma$ and 
\begin{equation}
    k_\Sigma^2 \sim \Sigma \sim \lambda' k_T^{d+2},
\end{equation}
where again $k_T \sim T^{-1/2}$. This gives $k_\Sigma \sim \sqrt{\lambda'} T^{-\frac{d+2}{4}}$. Hence, at long time there is an emergent length scale set by $k_\Sigma^{-1}$ which is parametrically longer than $k_T^{-1}$.

If we now add $\Sigma$ to the action and take a determinant we get
\begin{equation}
\det \begin{pmatrix}
\Sigma & i(i\omega+Dk^2)/2\\
i(-i\omega+Dk^2)/2 & D\kappa k^2
\end{pmatrix}= D^2k^4/4+\Sigma D\kappa k^2+\omega^2/4.
\end{equation}
From a calculation similar to that in section \ref{subsec:direct}, it follows that the coefficient of the ramp is modified to
\begin{equation}
    \log \text{coeff}(T)= - \frac{V}{(2\pi)^d}\int d^dk \ln\left( 1- \exp(-\sqrt{D^2k^4+4\Sigma D\kappa k^2}T)\right).
\end{equation}
or
\begin{equation}
\log \text{coeff}(T)=\frac{V}{(2\pi)^d}\int d^dk \sum_{j=1}^\infty \frac {\exp(-j\sqrt{D^2k^4+4\Sigma D\kappa k^2}T)}j.
\end{equation}
Since $k_\Sigma \ll k_T$, the integrand is always controlled by the diffusive scaling at large $k$ and it is smoothly cutoff for $|k|\gg k_T$. When the self-energy contribution can be ignored, we therefore find the usual result
\begin{equation}
    \log \text{coeff}_{\Sigma=0} \sim V k_T^d \sim V T^{-d/2}.
\end{equation}
The important question is whether the self-energy contribution significantly modifies this scaling. The self-energy contribution is only important for $|k|<k_\Sigma$, so restricting to that region of $k$-space and dropping the diffusive part, we find
\begin{equation}
    \log \text{coeff}_{\text{only $\Sigma$}} \sim V \int_0^{k_\Sigma} d|k| |k|^{d-1} \ln( 1- \exp(- \sqrt{4 D \kappa \Sigma} T |k|)).
\end{equation}
The typical size of the argument of the exponential is
\begin{equation}
    \sqrt{4 D \kappa \Sigma} T k_\Sigma \sim T \Sigma \sim T^{-d/2} \ll 1. 
\end{equation}
Hence, up to logarithmic corrections, the magnitude of the coefficient is just \begin{equation}
    \log \text{coeff}_{\text{only $\Sigma$}} \sim V k_\Sigma^d.
\end{equation}
This is always smaller than $V k_T^d$. 

A more complete picture of the integral is obtained by adding and subtracting the pure diffusive answer. Then, because argument of the exponential is close to zero for all $k$ where the $\Sigma$ term is important, we can approximate the logarithm as $\ln(1-e^{-u})\approx \ln u$ to give
\begin{equation}
    \log \text{coeff} = \log \text{coeff}_{\Sigma=0}- \frac{V}{(2\pi)^d} \int d^d k \ln \frac{\sqrt{D^2k^4+4\Sigma D\kappa k^2}}{D k^2}.
\end{equation}
The second term is indeed proportional to $V k_\Sigma^d$. The result is
\begin{equation}
    \log \text{coeff} = V \left(\frac{A_d}{T^{d/2}}-\frac{B_d}{T^{(d^2+2d)/4}} \right).
\end{equation}
In $d=1$, this reduces to
\begin{equation}
\label{eq:d1result}
    \log \text{coeff}_{d=1} = L \left(\frac{A_1}{T^{1/2}}-\frac{B_1}{T^{3/4}}\right).
\end{equation}
We conclude that the diffusive behavior dominates at late time, but there is a significant power-law correction to the diffusive behavior. Moreover, the correction is larger than what we found to leading order in perturbation theory.

\section{Extension to Floquet systems}
\label{sec:floquet}

In this section, we extend the previous theory to Floquet systems. There are several points to note. First, one can still define an SFF-like object in a Floquet system by restricting the time $T$ to be an integer multiple of the drive period. Chaotic Floquet systems are then expected to have an SFF which is given by random unitary ensemble. Such ensembles still give rise to a ramp, but with a different coefficient. One of our results is to point out that the hydro effective action can still correctly compute this coefficient. Second, one can still formulate a version of the CTP formalism even at infinite temperature, and one can use it compute the approach to the late time ramp, e.g. due to slow modes arising from a conserved charge. Third, one can also derive a hydro-like formula for the crossover from the Hamiltonian ramp to the Floquet ramp. We treat these three points in turn in the subsections below.

\subsection{The Circular Unitary Ensemble}

Consider first a Floquet system with no symmetry. If the Floquet dynamics is generated by a local time-dependent Hamiltonian $H(t)$ with period $2\pi/\Omega$, then the Floquet unitary is
\begin{equation}
    U_0 = \mathcal{T} \exp\left(-i \int_0^{2\pi/\Omega} ds H(s)\right),
\end{equation}
where $\mathcal{T}$ denotes time ordering and we integrate over a single period. The dynamics for $\Omega T/ 2\pi$ periods is given by 
\begin{equation}
    U(T) = (U_0)^{\Omega T/2 \pi}.
\end{equation}

Now, it is standard to define the Floquet Hamiltonian $H_F$ by \begin{equation}
    U_0 = e^{-i 2\pi H_F /\Omega}.
\end{equation}
Note that the energies of the Floquet Hamiltonian are only defined modulo $\Omega$. If the Floquet spectral form factor is defined as 
\begin{equation}
    \text{SFF} = \overline{|\text{Tr}[U(T)]|^2},
\end{equation}
then we see that the Floquet SFF evaluated at $\Omega T/2\pi$ periods is identical to the SFF of the Floquet Hamiltonian at time $T$. The important point about $H_F$ is that it is expected to be a very non-local object with no local conserved quantities.

We can thus obtain the predicted ramp coefficient using the DPT formalism as follows. Assuming all modes have non-vanishing lifetime as the system size goes to infinity, we can describe the SFF dynamics using just the spatial zero mode action
\begin{equation}
   I_{0} = - \int_0^T dt \phia \partial_t E_F,
\end{equation}
where now $E_F$ is the Floquet energy and $\Delta$ is proportional to the zero mode of $\phia$. The DPT path integral is
\begin{equation}
    Z_{\text{DPT}} = \int \frac{\mathcal{D}E_F \mathcal{D}\phia}{2\pi} e^{i I_0}.
\end{equation}
As in the Hamiltonian case, the action merely indicates that $E_F$ is conserved (modulo $\Omega$) and we are left with integrations over the zero frequency components of $\Delta$ and $E_F$. These are the relative time shift and the total Floquet energy. Using the same path integral normalization as in the Hamiltonian case, we find
\begin{equation}
    Z_{\text{DPT}} = \frac{T}{2\pi} \int dE_F.
\end{equation}
This is the correct result if the Floquet Hamiltonian is in the unitary class; otherwise there will be minor modifications to account for the symmetry structure. The final point is that the spectrum of $H_F$ is only defined modulo $\Omega$, so the energy integral is just $\Omega$. Hence, the spectral form factor is
\begin{equation}
    \text{SFF} = Z_{\text{DPT}} = \frac{\Omega T}{2\pi} = \text{number of periods},
\end{equation}
exactly as expected for the circular unitary ensemble (CUE). This immediately gives the late time result in~\cite{Friedman_2019}.

\subsection{The Floquet Chain With a Conserved Charge}

Now suppose the Floquet model has a U(1) symmetry described by diffusive dynamics at the quadratic level. We first define the analog of the CTP formalism. The main issue is just that, at infinite temperature, some of the CTP formalism degenerates and needs to be slightly modified. 

Consider the conserved charge \begin{equation}
    Q = \sum_r q_r
\end{equation}
and the associated lattice currents $j_{r,\hat{e}}$, where $\hat{e}$ denotes a direction on the $d$-dimensional hypercubic lattice. We couple these to the components of a background gauge field such that the time-dependent Hamiltonian becomes 
\begin{equation}
    H[A,t] = H(t) - \frac{\Omega}{2\pi} \sum_{r,\hat{e}} A_{r,\hat{e}}(t) j_{r,\hat{e}} - \frac{\Omega}{2\pi} \sum_r A_{r,0}(t) q_r .
\end{equation}
We assume for simplicity that $H(t)$ commutes with the total charge at each time $t$ (one can also consider Floquet models where only the integrated time evolution conserves the charge, but in this case we must slightly modify the definition of the coupling to $A$). A key point is that these gauge fields have unconventional units, since they are dimensionless.

Now, as in conventional CTP, we assume there is a long-wavelength description of 
\begin{equation}
    Z_{\text{CTP}} = \frac{1}{\text{Tr}(\mathbb{I})}\text{Tr}\left( U^\dagger(A_2) U(A_1)\right),
\end{equation}
where $U(A)$ denotes a product of time evolution steps as above. To quadratic order in fields and leading order in derivatives, the effective action is
\begin{equation}
    S_0 = \int d^d x dt \left[ i \chi_2 (\nabla \phi_a)^2 - \phi_a \left( \partial_t \rho - D \nabla^2 \rho \right)\right]
\end{equation}
with $\rho = \chi_1 \partial_t \varphi_r+\cdots$.
Based on this effective action, the analysis above in the context of energy diffusion will essentially carry over.

It is instructive to compare this result with the findings in~\cite{Friedman_2019}. That work derived the leading order diffusive approximation, what we have called the quadratic theory, from a different analysis method that is valid in the limit of large local dimension. Then it was shown that numerical data on spin chains also exhibited an approximate data collapse in the expected scaling variable $T/L^2$. Our analysis rerderives the same leading behavior from a different point of view. We also showed in Section ~\ref{sec:int_effects} that the leading effects of hydro interactions in the periodic time setting are subleading compared to the dominant diffusive behavior. This further justifies the validity of the diffusive form, which in \cite{Friedman_2019} is technically only justified at large local dimension and in our case is naively predicated on linear diffusion.

\subsection{Floquet Hydro With Weak Drive}

Let us now use hydrodynamics to study the relationship between the Hamiltonian ramp and the Floquet ramp. We now remove again the conserved $U(1)$ and consider a Floquet system with no exact conservation laws. One might think that hydro has no role to play here, apart from the zero mode calculation discussed above. However, there are two scenarios where such a driven system can be treated using hydrodynamics: a Taylor expansion in the strength of the driving force, or in the scenario where the period of the driving force is much longer than the equilibration time. In this paper we will only investigate the first of those possibilities.

We will eventually focus on the case of periodic driving, but for now we will be more general. Without the driving, we have energy conservation. In hydro language, this means we have a fluid time $\sigma$ in addition to the physical time $t$. It will be useful to define $\delta t=t-\sigma$ (in some other works, this is called $\epsilon$, but in this work $\epsilon$ is already defined as energy density). We will set $\phia=\delta t_1 -\delta t_2$.

Let's assume we have a time-dependent function $A(t)$ coupled to a (fast) mode $\phi$. At first we will consider an $A$ with period $T$, later we will consider the specific case where it has angular frequency  $\Omega=2\pi n/T$ with $n$ the number of periods. Now in addition to the usual hydro Lagrangian $L=\phia \partial_t E + \phi_a Q\phi_r+\dots$, we have an insertion of
\begin{equation}
    A_r(t(\sigma))\phi_a(\sigma)+A_a(t(\sigma))\phi_r(\sigma)\subset L
\end{equation} 

Remembering that $t$ is the physical time and $\sigma$ is hydrodynamic time, we can expand to leading order in $\delta t$. This becomes
\begin{equation}
    A_r(t(\sigma))\phi_a(\sigma)+A_a(t(\sigma))\phi_r(\sigma)\approx A(t) \phi_a(t)+A'(t)\phia\phi_r
\end{equation} 
At this point we integrate out the field $\phi$. We can shift $\phi$ to have an expectation value of zero. To leading order in the perturbation we will only be concerned with the two-point function. Inputting two-point functions $G_{ra}, G_{rr}$ from the microscopic theory we get 
\begin{equation}
   \int dt_1dt_2 G_{rr}(t_1-t_2)A'(t_1)\phia(t_1)A'(t_2)\phia(t_2)+\int dt_1dt_2G_{ra}(t_1-t_2)A(t_1)A'(t_2)\phia(t_2)\subset S_{\text{hydro}}.
    \label{eq:hydroFloquet}
\end{equation}  

One thing we need to remember is that at infinite temperature, $G_{ra}$ is zero, because when the density matrix $\rho_{\text{eq}}$ is proportional to the identity, all commutators vanish, $\tr( \rho_{\text{eq}}[\phi(t_1),\phi(t_2)])=0$. Perturbatively in the driving force, it makes sense to perform a Taylor expansion around infinite temperature. Expanding around $E=E_\infty$, the surviving term becomes $\int dt_1dt_2 \partial_E G_{ra}(t_2-t_2)A(t_1)A'(t_2)(E-E_\infty) \phia(t_2)$. Integrating out $t_1$ this becomes $ic\int dt_2 (E-E_\infty)\phia$. This can be thought of as a term allowing energy to relax towards $E_{\infty}$ as the system is driven. Including the normal $E\partial_t \phia$ term we get that the quadratic path integral is 
\begin{equation}
    \text{SFF}=\frac 1{1-e^{-cT}}.
\end{equation}
If we now assume an angular frequency for the drive $\Omega=2\pi n/T$, then there are $n$ saddle points corresponding to $\Delta=jT/n$ for $0\leq j<n$. So our formula becomes
\begin{equation}
    \text{SFF}=\frac 1{1-e^{-cT}}\frac{\Omega T}{2\pi},
    \label{eq:RoughFloquetSFF}
\end{equation}
where, again, $c$ is the rate of energy relaxation near infinite temperature. This is something we can calculate for a particular model.

To give a simple example, consider two SYK Hamiltonians~\cite{sachdev_sy_1993,kitaev_syk_2015,rosenhaus_syk_2016,maldacena_syk_2016} (defined and discussed in appendix \ref{app:SYK}) $H_1, H_2$ with their microscopic couplings correlated by some amount $r$, meaning the microscopic couplings for the same set fermions in $H_1$ and $H_2$ obey $\overline{J^1 J^2} = r \overline{(J^1)^2}$. We can alternate the two Hamiltonians, applying them for times times $T_1, T_2$ respectively, then repeat. The overall period is $\frac{2\pi}{\Omega} = T_1+T_2$. 

\begin{figure}
    \centering
    \includegraphics[scale=0.5]{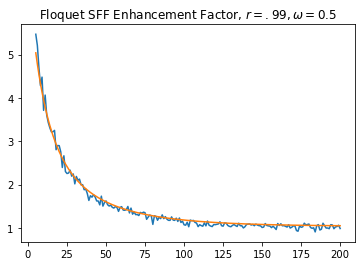}
    \includegraphics[scale=0.5]{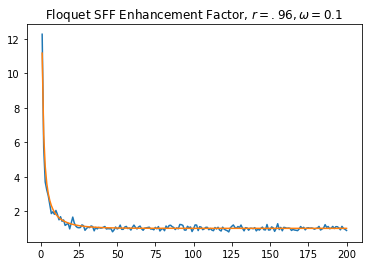}
    \caption{We numerically calculate the SFF for the driven system corresponding to two SYK Hamiltonians with microscopic couplings correlated at $r=.99/r=.96$ alternating back and forth. In both cases the numerical result for $\text{SFF}/\frac{\Omega T}{2\pi}$ (blue) lines up with the predicted $\frac 1{1-r^{-\Omega T/\pi}}$ (orange).}
    \label{fig:SYKdriven}
\end{figure}

If $T_1$ and $T_2$ are both longer than the Thouless time of $H_1$ and $H_2$ separately, then we can calculate $c$ fairly easily. Suppose the system starts at energy $E_i$ according to $H_1$. When the Hamiltonians flip, the system now has average energy $r E_i$ according to $H_2$. The system then thermalizes, and after the second flip it has energy $r^2 E_i$ according to $H_1$. Equating this reduction in energy to the accumulated decay rate over a single period, we get
\begin{equation}
e^{-2 \pi c / \Omega}=r^2.
\end{equation} 
For this system we thus have
\begin{equation}
    \text{SFF}=\frac 1{1-r^{-\Omega T/\pi}}\frac{\Omega T}{2\pi}.
\end{equation}
This is illustrated in figure \ref{fig:SYKdriven}.

\subsection{A Trace Formula for Floquet Systems}

The full hydrodynamic path integral with a time-dependent action \eqref{eq:hydroFloquet} is highly complicated and non-universal. There are, however, still some statements we can make. For instance, we have a generalization of the trace formula in equation \eqref{eq:bigAnswer}. We need to modify our definition of sectors so that they aren't weakly coupled sectors of the Hamiltonian with the same energy, but weakly coupled sectors of the unitary with the same phase. For instance, in undriven systems we can consider states with energy separated by $k \Omega$ to be in different sectors, since they have the same phase and will not mix. Driving then allows mixing between sectors with energy differing by a multiple of $\Omega$. The SFF (with general Dyson index restored) is then
\begin{equation}
    \textrm{SFF}= T \int \frac{d E_F}{\bbeta \pi} \tr \exp{M(E_F)T},
    \label{eq:floquetTransfer}
\end{equation}
where $M(E_F)$ is the transfer matrix connecting different sectors with the same quasi-energy $E_F$. In the limit of an undriven Hamiltonian with no internal structure, $M$ is diagonal in energy. The trace then just counts how many energies correspond to a given quasienergy. When we integrate this factor over the range of quasienergies, we get the range of energies, as expected.

In another limit, after enough driving all states with energies differing by a multiple of $\Omega$ will be fully mixed. The trace in Eq.~\eqref{eq:floquetTransfer} will then be unity, and the corresponding SFF is proportional to the number of periods up to a factor of the Dyson index,
\begin{equation}
    \textrm{SFF}=T \int \frac{dE_F}{\bbeta \pi}  = \frac{2}{\bbeta}\frac{\Omega T}{2 \pi}.
\end{equation}
The trace formula also generalizes equation \eqref{eq:RoughFloquetSFF} to general times given the explicit transition matrix $M$.

\begin{figure}
    \centering
    \includegraphics[scale=0.7]{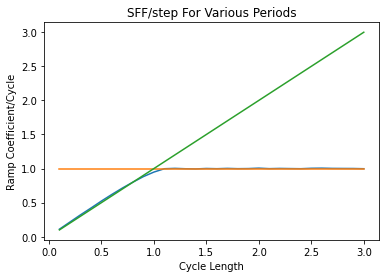}
    \caption{When the Floquet driving period is shorter than $2\pi/\textrm{Energy range}$, the ramp coefficient per period (blue) after many cycles is proportional to the period, being equal to the period times the energy range over $2\pi$ (graphed in green). When this quantity exceeds one, the coefficient per cycle is instead one (orange). This sharp crossover seen here in numerical data is also predicted by the theory.}
    \label{fig:sffpiecewise}
\end{figure}

Finally, note that the formula makes a sharp prediction at long times. If the range of energies in the unperturbed spectrum is $\delta E$, then we have a ramp coefficient per cycle of $\min(\delta E/\Omega, 1)$. Figure \ref{fig:sffpiecewise} is a graph illustrating this numerically.

\section{Discussion}

In this paper we developed a theory of the connected spectral form factor using tools from both RMT and hydrodynamics. This framework provides a number of key results, including formulas like equation \eqref{eq:bigAnswer} in cases with nearly conserved quantities, exemplified by equation \eqref{eq:returnCTP}. This allows us to rederive results previously only obtained for Floquet systems \cite{Friedman_2019,moudgalya2020spectral} using general hydrodynamic principles. We are also able to give new formulas that include nonlinear effects as in equation \eqref{eq:d1result}. Such nonlinearities are typically associated with long-time tails in hydrodynamics \cite{PhysRevA.4.2055}\cite{Kovtun_2003} and here we see them manifest in the spectral form factor. Our results shed light on how spatial locality in Hamiltonians interacts with ergodicity. Finally, while we focused on the simplest case of energy diffusion for simplicity, analogous results can be obtained in a wide variety of hydrodynamic theories. Quite generally, the Thouless time can be read off from the decay rates of the slowest hydrodynamic modes. And in a system without slow modes, the Thouless time should scale like the logarithm of the system size, since any mode with a system-size-independent decay rate will have a system size suppressed amplitude after logarithmic time.

One emerging lesson highlighted by our work is that quantum chaos should be viewed as a robust property of a (dynamical) phase of matter. In particular, the emergence of a pure random matrix ramp after the Thouless time is a feature that is stable to small perturbations. In fact, as we emphasized, the linear growth with time as well as the exact coefficient of the ramp are seemingly universal. What evidence is there for this? First, when considering deformations $H=H_0+g \delta H$, the derivative of the SFF with respect to $g$ is an expectation of an $a$-type variable and such expectations are suppressed by factors of $e^{-T/t_{\text{Th}}}$. This means the SFF is unaffected up to exponentially small corrections. Second, the addition of an ETH-obeying perturbation to the Hamiltonian corresponds to a stretching of the spectrum plus the addition of a random matrix. Hence, if the system had a linear ramp without this perturbation, it will also have one with the perturbation (see appendix \ref{app:folded}). Third, the basic phenomenon of the linear ramp comes from a symmetry breaking effect arising from the spontaneous breakdown of the relative time translation between the two SFF contours. Because this relative time translation symmetry cannot be explicitly broken by any time-independent Hamiltonian perturbation (i.e. without completely changing the problem), the corresponding symmetry broken phase should be both distinct from the unbroken phase and stable. 

When we glimpse different manifestations of quantum chaos like hydrodynamics and ETH connecting to the emergence of RMT, it suggests to us that a larger synthesis may be possible. Certainly there are many connections between chaos, random matrix statistics, and eigenstate thermalization, e.g. as reviewed in \cite{D_Alessio_2016}, as well as connections to notions of complexity, e.g.~\cite{Roberts_2017}. However, work remains to understand how all the different timescales obtained from various manifestations of chaos fit together, e.g.~\cite{dymarsky2018bound}. We hope to elaborate on these points in future work.

There are several issues that are still not fully understood, leaving room for further work. One is whether hydrodynamic methods can derive plateau behaviors in SFFs. Such a path integral derivation would need be be very unusual to reproduce the fact that plateau behavior is non-perturbative in the Heisenberg time, $T_{\textrm{Heisenberg}} \sim e^S$. Perhaps inspiration could be taken from other path-integral derivations of plateaus such as \cite{saad2019late,altland2020late,M_ller_2005}. Another question is the role of disorder. It seems that hydrodynamic SFFs naturally spit out values consistent with disorder averaging, despite there being no explicit disorder-averaging in the definition of the CTP formulation. Certainly for non-periodic times, the CTP formulation does not require disorder averaging to get correct real-time dynamics \cite{Blake_2018,glorioso2018lectures}. We need some disorder in order to make sense of the CTP action on the SFF contours, but this disorder can be small when the system size is large so that no intensive quantities, like transport parameters, are modified.

Finally, it is important to fully understand the possible effects of interactions in our modified CTP formalism. In the conventional CTP context, power counting indicates that interactions are irrelevant in the renormalization group sense. Interactions do generate novel effects not present in the Gaussian fixed point theory, but these effects can be captured in perturbation theory. The possibility of genuinely non-perturbative effects is not currently well understood. We showed that there are new effects arising from time periodicity, and carried out a partial resummation of diagrams to explore such effects. We found that they are subleading to the dominant quadratic behavior. However, it would be interesting to formulate a generalized renormalization group analysis to better understand the situation.

\section{Acknowledgements}

We thank Subhayan Sahu and Christopher White for helpful discussions throughout this process. This work is supported in part by the Simons Foundation via the It From Qubit Collaboration (B. S.) and by the Air Force Office of Scientific Research under award number FA9550-17-1-0180 (M.W.). M.W. is also supported by the Joint Quantum Institute.

\appendix
\section{Appendix: Review of the SSS Wormhole Solution}
\label{app:SSS}

In \cite{saad2019semiclassical} the authors (SSS) evaluate the ramp contribution exactly for the SYK model (see Appendix \ref{app:SYK}). They start by noting that the SFF is a partition function on two contours. In particular, if we denote the collective fields of a single copy of the system by $\Psi$, the SFF is given by a path integral on two copies of the system with periodic time coordinates,
\begin{equation}
    \sff(T,f(E)=1)=\int \mathcal D\Psi_1D\Psi_2 \exp\left (i \int d^d x\int_0^T dt \{ L[\Psi_1]-L[\Psi_2]\}\right)
\end{equation}
The essential insight in their paper is that for systems with many degrees of freedom, like the SYK model at large $N$, this path integral can be evaluated by saddle-point methods and that a non-trivial family of saddle points give the ramp. These are thermofield double solutions at inverse temperature $\beta_{\text{aux}}$, suitably adjusted using images to account for the different boundary conditions, that correlate the two contours. At large $T$ such solutions always approximately solve the SFF two contour equations of motion because they solve the equations of motion on a $e^{-\beta_{\text{aux}}H}e^{iHT}e^{-iHT}$ contour (where the time evolutions trivially cancel) and the contours are identical in the `bulk' of the forward and backward legs. Implicitly, we are appealing to the forgetfulness of chaotic systems, which here means that the solutions are exponentially insensitive to the boundary conditions.


\section{Appendix: Review of The SYK Model}
\label{app:SYK}

The Sachdev-Ye-Kitaev (SYK) model is a disordered 0+1d system made of Majorana fermions with $q$-body interactions ($q$ is even)~\cite{sachdev_sy_1993,kitaev_syk_2015,rosenhaus_syk_2016,maldacena_syk_2016}. The SYK Hamiltonian is given by
\begin{equation}
    H[\psi] = i^{q/2}\sum_{1 \leq j_1 <...< j_q \leq N} J_{j_1j_2...j_q}\psi^{j_1}\psi^{j_2}...\psi^{j_q},
    \label{eq:SYKHamiltonian}
\end{equation}
where $\psi^i, i=1,...,N$ represents the Majorana fermions and satisfy the anticommutation relation $\{\psi^i,\psi^j\}=\delta_{ij}$, and each $J_{j_1...j_q}$ is a Gaussian variable with mean zero and variance $\langle J_{j_1...j_q}^2 \rangle=\frac{J^2 (q-1)!}{N^{q-1}}$. 

It is often convenient to perform a series of exact manipulations on Hamiltonian \eqref{eq:SYKHamiltonian} to get a mean-field Lagrangian description of the SYK model in terms of bilocal variables consisting of a Green's function $G$ and self-energy $\Sigma$. In particular, one can write an expression for the imaginary temperature partition function of the SYK Model as
\begin{equation}
Z(iT)=\int DG D\Sigma \exp N \Big[ \frac12 \Tr \log( \partial_t - i\Sigma)\\
    + \frac 12\int dt_1 dt_2 (i \Sigma(t_1,t_2) G(t_1,t_2) - \frac{J^2 }{q} G^q(t_1,t_2) ) \Big].
\end{equation}
The SFF can be thought of as a partition function of a doubled system living on two contours, with one contour running forward in time (corresponding to $e^{-i H T}$ in the SFF) and one contour running backward in time (corresponding to $e^{i H T}$ in the SFF). Generalizing the result for $Z(iT)$, one can write the SFF as
\begin{equation}
\begin{split}
   \sff(T,\beta=0) = \int DG D\Sigma\exp N \Big[ \frac12 \Tr \log( \partial_t - i \hat\Sigma)\\
    + \frac{1}2 \sum_{\alpha,\beta=1,2}\int dt_1 dt_2 (i \Sigma_{\alpha \beta} G_{\alpha\beta} - \frac{J^2 }{q}  (-1)^{\alpha+\beta} G_{\alpha\beta}^q ) \Big] , 
\label{eq:Action}
\end{split}
\end{equation}
where a hat above a variable signals a matrix representation, $(\hat \Sigma)_{\alpha\beta} \equiv \Sigma_{\alpha\beta}$. Because of the antiperiodic boundary conditions on the fermions, both $G_{\alpha \beta}$ and $\Sigma_{\alpha \beta}$ are antiperiodic under time shifts by $T$. Note also that the measures $DG$ and $D\Sigma$ each integrate over the space of two-index functions of two variables.

\section{Appendix: CTP Formulation of Hydrodynamics}
\label{app:CTP}

The Closed Time Path (CTP) formalism \cite{glorioso2018lectures,crossley2017effective} is an effective theory of hydrodynamics on the Schwinger-Keldysh contour. In \cite{Chen_Lin_2019} a simplified version describing just energy diffusion is used to derive long-time-tails for two-point functions in hydrodynamics. We will largely follow their conventions. 
One starts with a partition function
\begin{equation}
    Z_{CTP}[A_1^\mu(t,x),A_2^\mu(t,x)]=\frac {1}{\tr e^{-\beta H}} \tr \mathcal P e^{-\beta H}\exp(-i H T+\int dx dt A_1^\mu J_{1\mu})\exp(i H T-\int dx dt A_2^\mu J_{2\mu}),
\end{equation}
where $A$ can be thought of as an external gauge field coupling to the conserved currents.

We express $Z_{CTP}$ as $e^{I[A_1,A_2]}$, where $I$ is a nonlocal action. The main assumption is that after `integrating in' slow modes the action will become local. In a standard hydrodynamic system, the only slow modes correspond to conservation laws and these modes are brought in to enforce those laws. We have
\begin{equation}
    \exp(I[A^\mu_1,A^\mu_2])=\int \mathcal D \phi_1\mathcal D \phi_2\exp(i\int dtdx L[B^\mu_1=A^\mu_1+\partial^\mu \phi_1,B^\mu_2=A^\mu_2+\partial^\mu \phi_2)
    \label{eq:CTPAction}
\end{equation}
Where $L$ is a (generally complex) action functional. 

As equation \eqref{eq:CTPAction} makes manifest, the action doesn't depend on the $\phi$s except through the modified gauge fields $A_\mu+\partial_\mu \phi$. If we change variables to
\begin{equation}
\begin{split}
    \phi_r=\frac{\phi_1+\phi_2}{2}\\
    \phi_a=\phi_1-\phi_2
\end{split}
\end{equation}
we can derive additional identities. For instance unitary implies that 
\begin{equation}\left\langle \prod_i \phi_i(t_i)\phi_a(0)\right\rangle =0\end{equation}
will always be zero whenever $\forall_i t_i<0$. In other words, if the chronologically latest insertion is $a$-type, the expectation value is zero. This theorem is called the last time theorem, and is explored in detail in \cite{gao2018ghostbusters}. This, in turn, can be used to derive the fact that all terms in $L$ have at least one factor of an $a$ field. It is also worth explicitly noting the implications of the last time theorem for two point functions. We also always have
\begin{equation}
\begin{split}
    G_{aa}(t)=0,\\
    G_{ra}(t)=\theta_+(t) f(t).
\end{split}
\end{equation}
Additional constraints on CTP Lagrangians derived in other works include the fact all factors of the $r$ variables come with at least one time derivative and the KMS condition that $S[\phi_1(t),\phi_2(t)]=S[\phi_2(i\beta-t),\phi_1(-t)]$.

It is also worth noting that we can formulate in terms of slightly different variables, replacing $\partial_t \phi_r$ with the energy density $\epsilon$. For the theory in \cite{Chen_Lin_2019} used in the main text, at the Gaussian level $\epsilon$ is equal to $c \beta^{-1} \partial_t\phi_r$ plus higher derivative corrections. The general formula for $\epsilon$ is obtained by differentiating the action with respect to $A^0_a$, since it couples to $\epsilon$. For one conserved quantity at leading order in derivatives, the most general quadratic Lagrangian consistent with the requirements of derivatives and KMS symmetry is
\begin{equation}
  L=-\phi_a\left(\partial_t\epsilon-D\nabla^2\epsilon\right)+i\beta^{-2}\kappa(\nabla \phi_a)^2.
\end{equation}

To make the physics more transparent, it is useful to introduce an auxiliary variable $F(x,t)$. We can rewrite our action as
\begin{equation}
    L=-\phi_a\left(\partial_t\epsilon-D\nabla^2\epsilon - F\right)+i\frac{1}{4\beta^{-2}\kappa} F \nabla^{-2} F.
\end{equation}
The action is now linear in $\phi_a$, meaning that $\phi_a$ serves as a Lagrange multiplier enforcing the stochastic partial differential equation
\begin{equation}
    \partial_t\epsilon-D\partial_x^2\epsilon+F=0,
\end{equation}
where $F$ is now interpreted as a fluctuating force. This then leads to the probability distribution of energy modes discussed in the main text.

\section{Appendix: Folded Spectra}
\label{app:folded}

Here we comment on the case of stretched and folded spectra. To start with, consider a random Hamiltonian $H'=\fstretch(H)$, where $H$ is a random matrix chosen from distribution \eqref{eq:RMTpdf} and $\fstretch$ is a smooth function with everywhere positive derivative. The quantum mechanics of such deformations have been considered recently in \cite{Gross_2020}. Another motivation to study folded spectra comes from the eigenstate thermalization hypothesis (ETH). ETH asserts that any local observable $ O$ can be written as a sum of a smooth function of energy $f_{O}(H)$ (related to the microcanonical expectation value) and a random-like erratic part $R$~\cite{PhysRevE.50.888,PhysRevA.43.2046}. Under this hypothesis, the SFF of a Hamiltonian perturbed by a local operator is then equivalent to the SFF of a stretched spectrum plus a random matrix, $H' = H + \epsilon O \sim H + \epsilon f_O(H) + R$.

Studies of the SFFs of folded systems are common~\cite{Gharibyan_2018,Gubin_2012,Vahedi_2016,phdthesis}. One reason is that a folding procedure (often called `unfolding') can be used to get semicircle statistics out of other level distributions in order to more easily compare numerical results with RMT. In this section we show analytically that non-singular folds indeed leave ramps invariant. For a comparison of folding versus filters as a way to look at parts of the spectrum see \cite{Nosaka_2018}.

Returning to $H' =\fstretch(H)$, there is generically no $V'$ such that $H'$ is distributed according to \eqref{eq:RMTpdf}. Rather, the pdf for $H'$ is given by
\begin{equation}dP=\frac 1 {\mathcal Z}\prod_{i<j} |\fstretch^{-1}(\lambda_i)-\fstretch^{-1}(\lambda_j)|^{\boldsymbol \beta} \prod_i e^{-V(\fstretch^{-1}(\lambda_i))+\log \fstretch^{-1'}(\lambda_i)}.
\label{eq:RMTpdfS}
\end{equation}
Nonetheless, the spectral statistics of $H'$ are very similar to those given by \eqref{eq:RMTpdf}. 
This is because nearby eigenvalues still repel with repulsion term $(\fstretch^{-1}(E'_1)-\fstretch^{-1}(E'_2))^\bbeta$ which is roughly proportional to $(E'_1-E'_2)^\bbeta$. As such, the ramp still exists with coefficient given by \eqref{eq:RMTAnswer}. 

Another way to see this is to consider a Gaussian filter function. If the variance $\sigma$ in the filter function is small compared to the scale of variation in $f_{\text{stretch}}$\footnote{For example, if $f_{\text{stretch}}$ is a slowly varying function of the energy density.}, then for the small window around $\bar E$, the stretching simply rescales all the differences between eigenvalues by $f_{\text{stretch}}'(\bar E)$, which is a trivial change. The effect on SFFs with broader filter functions can be obtained by integrating over $\bar E$.

\begin{figure}
\begin{tabular}{cc}
\includegraphics[scale=0.5]{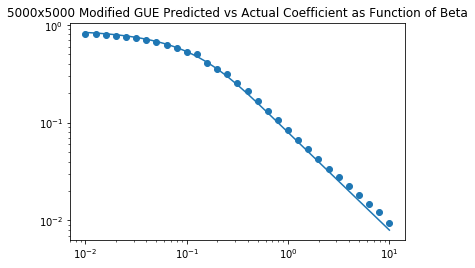}&\includegraphics[scale=0.5]{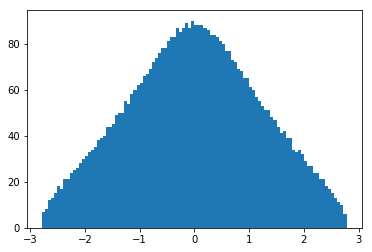}\\
\includegraphics[scale=0.5]{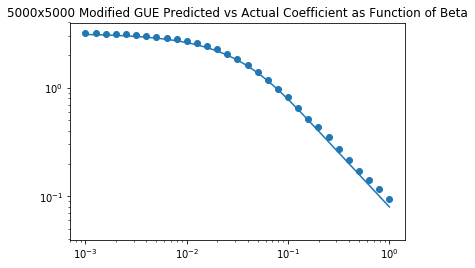}&\includegraphics[scale=0.5]{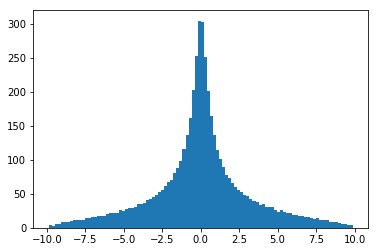}
\end{tabular}
\caption{Ramp coefficient plots and spectral densities for $\fstretch(E)=E+0.1E^3$ and $\fstretch(E)=E+E^3$.}
\label{fig:stretch}
\end{figure}

Figure \ref{fig:stretch} shows coefficient plots of  5000 by 5000 GUE matrices after transformations $\fstretch(E)=E+0.1E^3$ and $\fstretch(E)=E+E^3$, accompanied by histograms of their spectral density

The next natural is question to ask is what happens when we choose a function $\fstretch$ which doubles back on itself, for instance $\fstretch(E)=E-E^3$. In these cases we can have multiple `species' of eigenvalues near $E'$, corresponding to which branch of $\fstretch^{-1}$ the original $E$ lies on. There is almost no repulsion between different species of eigenvalue, so the ramp part of the SFF is given by
\begin{equation}
\int dE f^2(E) \frac T {\pi\bbeta} (\# \textrm{ of species at }E)
\label{eq:species}
\end{equation}

\begin{figure}
\begin{tabular}{cc}
\includegraphics[scale=0.5]{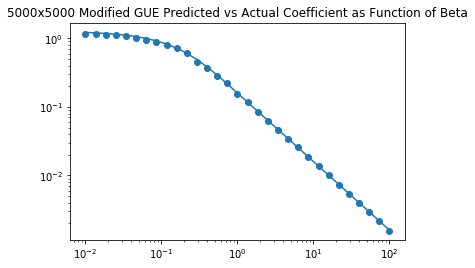}&\includegraphics[scale=0.5]{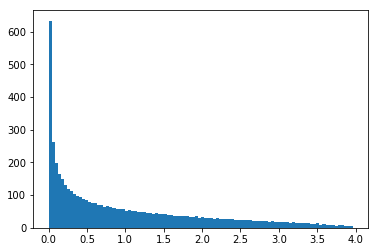}\\
\includegraphics[scale=0.5]{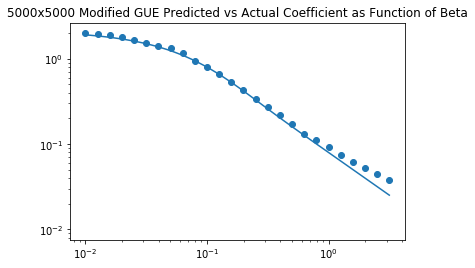}&\includegraphics[scale=0.5]{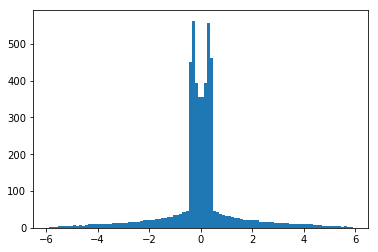}
\end{tabular}
\caption{Ramp coefficient plots and spectral densities for $\fstretch(E)=E^2$ and $\fstretch(E)=E-E^3$.}
\label{fig:stretchSpecies}
\end{figure}

Figure \ref{fig:stretchSpecies} shows the matching between \eqref{eq:species} and numerical experiment for $\fstretch(E)=E^2$ and $\fstretch(E)=E-E^3$.
One question that remains open is what the behavior is like near the turning points, characterized by $\frac{d}{dE}\fstretch(E)=0$, where the repulsion term becomes singular. Might there be a strong enough contribution to change the overall behavior?

\bibliography{rampCoefficient.bib,ApproximateSymmetries.bib}

\end{document}